\pdfoutput=1

\documentclass[10pt,a4paper,twocolumn]{article} % onecolumn or twocolumn

\usepackage[version=3]{mhchem} % Package for chemical equation typesetting
\usepackage{siunitx} % Provides the \SI{}{} and \si{} command for typesetting SI units
\usepackage{braket}
\usepackage{graphicx} % Required for the inclusion of images
\usepackage[numbers,sort&compress]{natbib}
\usepackage{amsfonts,amsmath,amsthm,amssymb,amstext}% erweiterte Funktionalität für Formeln (Pakete der American Mathematical Society)
\usepackage[utf8]{inputenc}% Text-Encoding festlegen. Mit utf8 oder utf8x funktionieren 
\usepackage[english]{babel}
\usepackage{microtype} % optimiert das typographische Erscheinungsbild
\usepackage{enumitem}% erlaubt Listen einfacher zu formatieren (bietet nosep für kompakte Listen)
\usepackage{ctable} % erlaubt hübsche Tabellen über mehrere Seiten, beinhaltet booktabs (\toprule, \midrule, ...)
\usepackage{xcolor} % ermöglicht farbigen Text ({\color{red} ...})
\usepackage{mathrsfs}
\usepackage{mdframed} % frames around something [linecolor=xxx,backgroundcolor=xxx, linewidth=xxpt,topline(rightline, leftline)=false]
\usepackage{tabu}
\usepackage{booktabs} %toprule,midrule,bootomrule
\usepackage{caption}
\usepackage{graphicx}%% erlaubt es Bilddateien einzubinden (ctable graphicx intern auch. Trotzdem ist es sinnvoll graphicx expilizt zu laden,sonst entstehen schwehr verständliche Fehler, wenn ctable entfernt wird)
\usepackage{float}% ermöglicht Bilder und Tabellen am eingegebenen Ort zu platzieren ([H])
\usepackage{subfig} % ermöglicht Unter-Bilder in einer figure-Umgebung
\graphicspath{{img/}}% Grafik-Dateien werden in den folgenden Ordnern gesucht
\DeclareGraphicsExtensions{.pdf,.png,.jpg}% Grafikdateien haben die folgenden Endungen (höchste Priorität zu erst)
\usepackage[
a4paper,
total={16cm,26cm},          % Breite und Höhe des Inhalt-Bereichs
top=20mm, left=30mm,        % Ränder oben und links
headsep=10mm,               % Abstand des unteren Rands der Kopfzeile vom oberen Rand des Inhalts
footskip=10mm               % Abstand des unteren des Inhalts zum oberen Rand der Fusszeile
]{geometry}

\usepackage[
pdftex,                        % wir verwenden pdftex/pdflatex
bookmarks=true,                % wir wollen auch im PDF-Reader ein Inhaltsverzeichnis
bookmarksdepth=3,              % das Inhaltsverzeichnis soll 3 Tiefen enthalten
colorlinks=true,               % Linktexte sollen Farbig sein
linkcolor=black,               % Links innerhalb des Dokuments bleiben schwarz
citecolor=black,               % Links zu Quellenangaben bleiben ebenfalls schwarz
urlcolor=blue,                 % URL-Linktexte sollen blau dargestellt werden
]{hyperref}

\usepackage{wrapfig}

\usepackage[english, capitalise]{cleveref}
\usepackage{adjustbox}
\usepackage{pdflscape}

\title{AI-supported Degradation Study of Carbon-based Perovskite Solar Cells, Learning the Device Physics of Perovskite Solar Cells: A Drift–Diffusion-Guided Autoencoder Approach}  % Title or \title{\vspace{-xcm}Title} to move upwards

\author{Oliver Zbinden, Sharun Parayil Shaji, Wolfgang Tress}

\date{}

\begin{document}
%\clearpage
\maketitle
%\thispagestyle{empty}	
%\maketitle % Insert the title, author and date

\begin{abstract}

Carbon-electrode-based PSC devices are stressed under 1 Sun equivalent illumination in a stability setup, and different scan-speed dependent current-voltage (J-V) curves are measured during aging. The collected data is used to estimate several physical parameters that contain information about charge transport and recombination using Machine Learning (ML), which allows for in situ tracking of possible signs of degradation. These results are compared to what can be classically interpreted by analysing changes in J-V curves, and the evolution of the predicted parameters is studied. The predictions are then used to simulate a digital twin of the measured devices, and their physical implications and the differences between measurements and devices are discussed.
\end{abstract}

\section{Introduction}

Since their invention \cite{kojima}, perovskite solar cells (PSCs) have experienced an impressive increase in efficiency from 3.8\% to almost 27\% \cite{nrel}, which is only 0.3\% behind silicon solar cells. The record for carbon-based PSCs recently reached 23.2\% \cite{carbonrecord}. Still, some problems have to be overcome, most importantly, degradation. Despite huge progress in understanding the underlying mechanisms \citep{leguy,conings,bryant,li_light,aristidou,min-cheol,dongxu,girolamo,jacobs} and improvements \citep{seongkang,jaesung,grancini,sominpark,boyd,zhao}, these problems are still not fully solved.
Regarding stability and degradation, most studies using Artificial Intelligence (AI) focused on overall device performance, such as predicting the evolution of power conversion efficiency (PCE) over time \cite{odabasi,zhaojie,hartono}, or the time it takes for PCE to drop to $80\%$ of its initial value, $\rm T_{80}$ \citep{graniero,kouroudis,mammeri,dunlap}. This is very helpful for predicting the lifetime of a device and for quality control. However, these applications do not directly contribute to the improvement of PSCs or identify sources of PCE loss.
ML has recently been shown to provide reliable estimations of device parameters based on J-V curve measurements \cite{AE_paper} This approach allows for the estimation of parameters of devices, for example, to improve the accuracy of measurement matching simulation.

In this paper it is demonstrated how ML can be used to track parameter changes that cannot be directly measured during stressing the devices without interruptions, such as transfer to different measurement setups. Two devices from \textit{Solaronix} \cite{solaronix} are kept in a stability setup inside inert atmosphere with constant illumination for a total of 23 days, one device under maximum power point (MPP) tracking, and the other device under open circuit voltage ($\rm V_{OC}$) tracking. Two additional unstressed devices were kept as references. Stressing devices in such a way is likely to cause degradation \cite{nie,domanski}. The tracking is interrupted periodically and J-V curves are measured every day. With measured J-V curves, ML is used to estimate several physical parameters, allowing for tracking their evolution over time. 1D drift-diffusion (DD) device simulation allows a comparison between the measured J-V curves and a model with the obtained parameter estimates. The evolutionary trend of each parameter series is discussed and analyzed, based on the influence of possible degradation effects.

%$\rm J_{SC}$ decreases when device is put under 1 Sun illum at $\rm V_{OC}$ \cite{nie}

\section{Methods}

The simulation and ML architecture are adopted from \cite{AE_paper}. For this study, the same training protocol is used.

\subsection{Device Simulation}

\begin{figure}[t]
    \centering
    \includegraphics[width=1\linewidth]{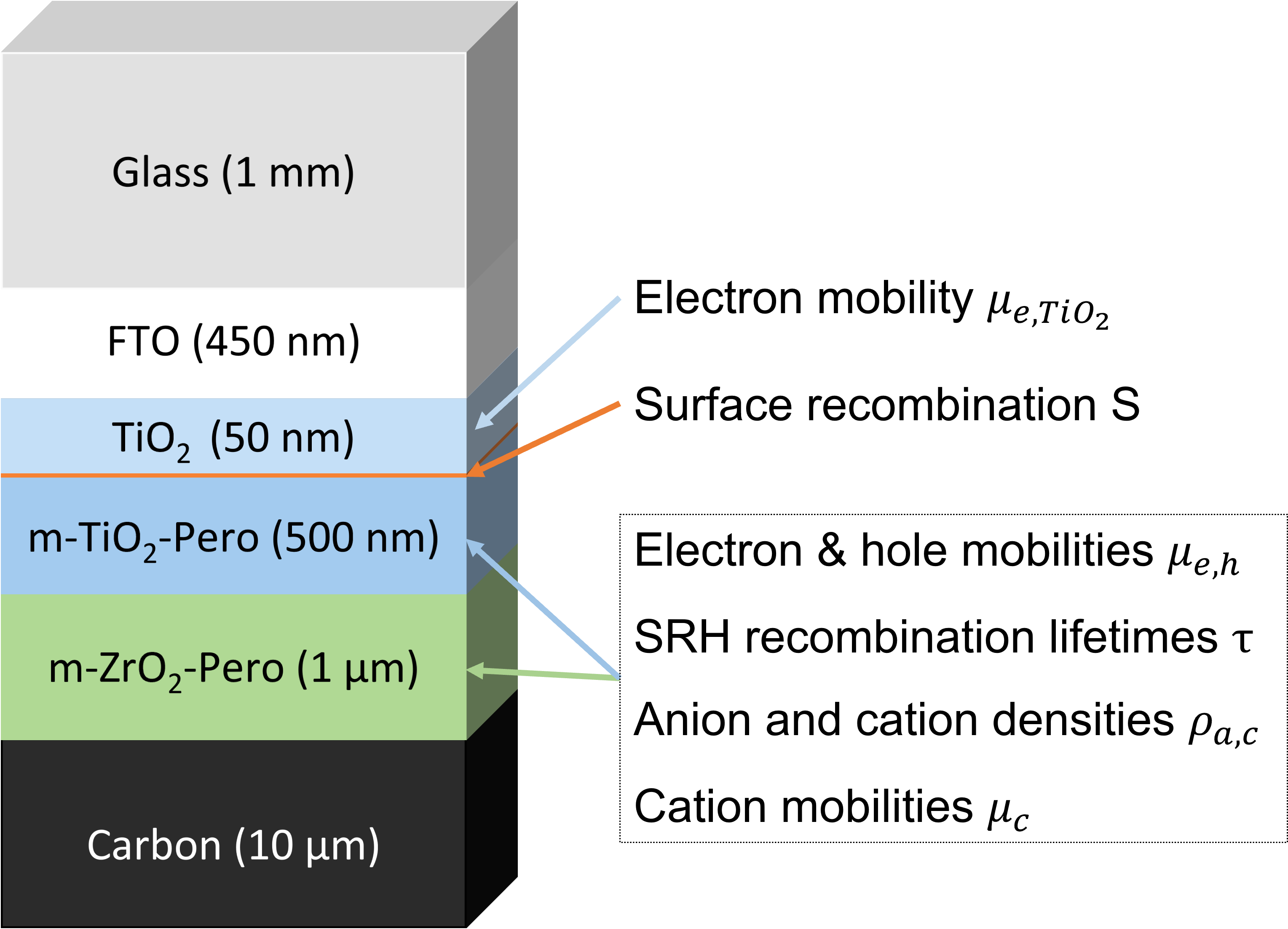}
    \caption{Illustration of the device stack of the mesoporous PSCs used in experiment and simulation. The arrows indicate the layers where parameters were varied.}
    \label{fig:devstackdegr}
\end{figure}

Starting with a carbon electrode-based mesoporous metal-oxide-perovskite based \cite{kerremans,bogachuk_comparison_2021}, with an FTO/TiO$_2$/m-TiO$_2$-MAPI/m-ZrO$_2$-MAPI/Carbon structure (Figure \ref{fig:devstackdegr}), several material parameters are varied randomly, each of them independently.
The varied parameters are the surface recombination velocity between the mesoporous perovskite-TiO$_2$ and TiO$_2$ layer $\rm S_{e,\, top},\, S_{h,\, top}$, in the mesoporous perovskite-metal oxide layers electron and hole mobilities $\rm \mu_e,\, \mu_h$, electron and hole recombination lifetimes $\rm \tau_e,\, \tau_h$, anion and cation densities $\rm \rho_A,\, \rho_C$, electron mobility in TiO$_2$ $\rm \mu_{TiO_2}$, and cation mobilities $\rm \mu_C$. These parameters are selected based on their influence on the device, in order to cover various phenomena, such as changes related to charge transport vs. charge recombination, or electronic vs. ionic effects. These parameters are expected to show some sort of variability between individual devices, influenced by different conditions during fabrication, or caused by material changes during degradation \cite{dunfield_deg_incr_ion,zhang_degradation_2022,kim_degrad_chargemob}. To reduce the complexity of the DD model, we set $\rm\mu_e=\mu_h =: \mu$, $\rm S_{e,\,top}=S_{h,\,top} =: S$, $\rm\tau_e=\tau_h=:\tau$, and $\rm\rho_A=\rho_C=:\rho_{Ion}$. Note especially that the parameters in the m-TiO$_2$-MAPI and m-ZrO$_2$-MAPI layers are assumed to be the same regarding charge transport and recombination, they only differ in their optical and light absorption properties.
Because most reported parameters are determined for pure perovskite layers (for example in \cite{neukom_consistent_2019}), there is only little information for mesoporous perovskite-metal-oxide materials \cite{kerremans}. Therefore, the ranges within the parameters are varied are relatively wide, but still within physically reasonable boundaries, as listed in Table \ref{tab:paramsweepsdegstudy}.

%\begin{table}[t]
%    \centering
%    \caption{Literature-oriented parameter range for the different quantities, based on physically reasonable boundaries. Cation mobilities $\rm\mu_C$ were only varied in the transient case.}
%    \label{tab:paramsweepsdegstudy}
%    \begin{tabular}[htbp]{lrrl}
%        \toprule
%        Param. & Minimum & Maximum & Units\\
%        \midrule
%        $S$ & $10^{-3}$ & $100$ & $\rm cms^{-1}$ \\
%        $\rm \mu$ & $10^{-2}$ & $10^3$ & $\rm cm^2V^{-1}s^{-1}$ \\
%        $\tau$ & 1 & $10^4$ & $\rm ns$ \\
%        $\mu_{TiO_2}$ & $2.5\cdot10^{-5}$ & 1 & $\rm cm^2V^{-1}s^{-1}$ \\
%        $\rm\rho_{Ion}$ & $10^{14}$ & $10^{20}$ & $\rm cm^{-3}$\\
%        $\rm\mu_C$ & $10^{-15}$ & $10^{-4}$ & $\rm cm^2V^{-1}s^{-1}$ \\
%        \bottomrule
%\end{tabular}
%\end{table}

\begin{table}[t]
    \centering
    \caption{Literature-oriented parameter range for the different quantities, based on physically reasonable boundaries. Cation mobilities $\rm\mu_C$ were only varied in the transient case.}
    \label{tab:paramsweepsdegstudy}
    \begin{tabular}[htbp]{lrrl}
        \toprule
        Param. & Minimum & Maximum & Units\\
        \midrule
        $S$ & $10^{-3}$ & $100$ & $\rm cm/s$ \\
        $\rm \mu$ & $10^{-2}$ & $10^3$ & $\rm cm^2/Vs$ \\
        $\tau$ & 1 & $10^4$ & $\rm ns$ \\
        $\mu_{TiO_2}$ & $2.5\cdot10^{-5}$ & 1 & $\rm cm^2/Vs$ \\
        $\rm\rho_{Ion}$ & $10^{14}$ & $10^{20}$ & $\rm cm^{-3}$\\
        $\rm\mu_C$ & $10^{-15}$ & $10^{-4}$ & $\rm cm^2/Vs$ \\
        \bottomrule
\end{tabular}
\end{table}

This wide range of varied parameters leads to simulated J-V curves that cover a space reaching from a sample that can be considered as completely degraded, up to a hypothetical device that outperforms the measured devices. This introduces a good amount of generalizability, which is needed for ML training. To detect the influence of mobile ions, different scan speeds are simulated ($0.005$, $0.05$, $0.5$, $5$, $50$, $100$) V/s in decreasing order, first in forward, then in reverse direction ($0$ V - $1.2$ V - $0$ V).

\subsection{ML Model Description}

The encoder part of the AE consists of two convolutional and three dense layers, the decoder is mirror-symmetric. To find the parameters of interest, the loss function consists of two terms. The first term compares the input, the simulated J-V curves, to the AE output, the reconstructed J-V curves. The second additional term in the loss function, which can be seen as a condition, compares the true parameter values of the training data known from simulation, both terms are weighted equally. The second term in the loss function is crucial for two reasons. First, a normal AE undercomplete compresses the input in a way that retains necessary information to reconstruct the input, but in a "black-box" way. It could not be verified that the latent parameters indeed contain information of the parameters varied in simulation, and, if so, the function that maps the latent parameters to the physical parameters would be unknown. Second, assuming that the simulation parameters are mapped to the latent space independently without convoluting parameters, the order of parameters in the latent vector would be unknown, with $6!=720$ possible combinations for the given task with 6 differently varied parameters.

The input of the AE is a 1D array with length 732 that consists of the simulated or measured current values in increasing order of scan speeds starting with the forward direction, interpolated to $\rm 0.02$ V steps. The current arrays are normalised with the global minimum and maximum from the simulated data. The six varied parameters, $\rm S$, $\rm \mu$, $\rm \tau$, $\rm \mu_{TiO_2}$, $\rm \mu_C$, and $\rm \rho_{Ion}$ are collected in a 1D array for each simulated device, and log-normalised with the respective global minimum and maximum known from simulation. In total, there are $50'000$ simulated devices that are split into fractions of $70:15:15$ for training, validation, and test sets. Training is performed in a $10-$fold cross-validation.

After training is completed, the encoder and decoder parts can be used independently. The decoder can act as a fast surrogate device simulator, but this part of the AE is discarded since, for parameter estimation, only the encoder is needed. 

%\begin{table}[t]
%    \centering
%    \caption{$\rm R^2$ and $f_{0.2}$ (fraction of estimated parameter values with an absolute residual smaller than 0.2 for the log-normalised parameters) as a quality measure. \textcolor{red}{this was not really part of the degradation study. should i keep it here or would it be better to move it to the results section?}}
%    \label{Table_1}
%    \begin{tabular}[htbp]{@{}lcc@{}}
%		\toprule
%		Parameter & R$^2$ & $f_{0.2}$ \\
%		\midrule 
%		$S$ &  $0.500$ & $0.703$  \\
%        $\rm \mu$ & $0.984$ & $0.993$ \\
%        $\tau$ & $0.888$ & $0.942$ \\
%        $\mu_{TiO_2}$ & $0.351$ & $0.666$ \\
%        $\rm\rho_{Ion}$ & $0.859$ & $0.941$ \\
%        $\rm\mu_C$ & $0.961$ & $0.985$ \\
%        %\midrule
%        %$J$ & $0.988$ & - \\
%\bottomrule
%\end{tabular}
%\end{table}

\subsection{Device Measurement}

J-V measurements were performed using a PAIOS system (Fluxim AG) under simulated AM1.5G illumination (Oriel Sol3A Class ABB, model 94011A). Encapsulated devices were preconditioned under one sun illumination at open-circuit voltage for 5 minutes prior to the J-V measurements. Subsequently, J-V curves were recorded sequentially at scan rates of 100, 50, 5, 0.5, 0.05, and 0.005 V/s, in that order, without intermediate light soaking or voltage preconditioning. For each scan rate, the measurement was performed first in the forward direction (from short-circuit to open-circuit voltage), followed by the reverse direction. All measurements were carried out in ambient atmosphere.

Light-intensity dependent J-V measurements at scan speed of 0.036 V/s were also performed using PAIOS by  varying the illumination (White LED) intensity from aprox. 2 to 0.02 sun eqivalent intensity  to analyze recombination behaviour and extract ideality factors.

Stressing of the cells was done under MPP and $\rm V_{oc}$ with white led (Colour temperature = 5000K) illumination of one sun equivalent by employing Maximum power point tracker (MP0205M24 from LPVO) . During the stressing we measured the J-V as mentioned above using the potentiostat EmStast-4s HR from PalmSense.

\section{Results}

\begin{figure}[t]
    \centering
    \includegraphics[width=1\linewidth]{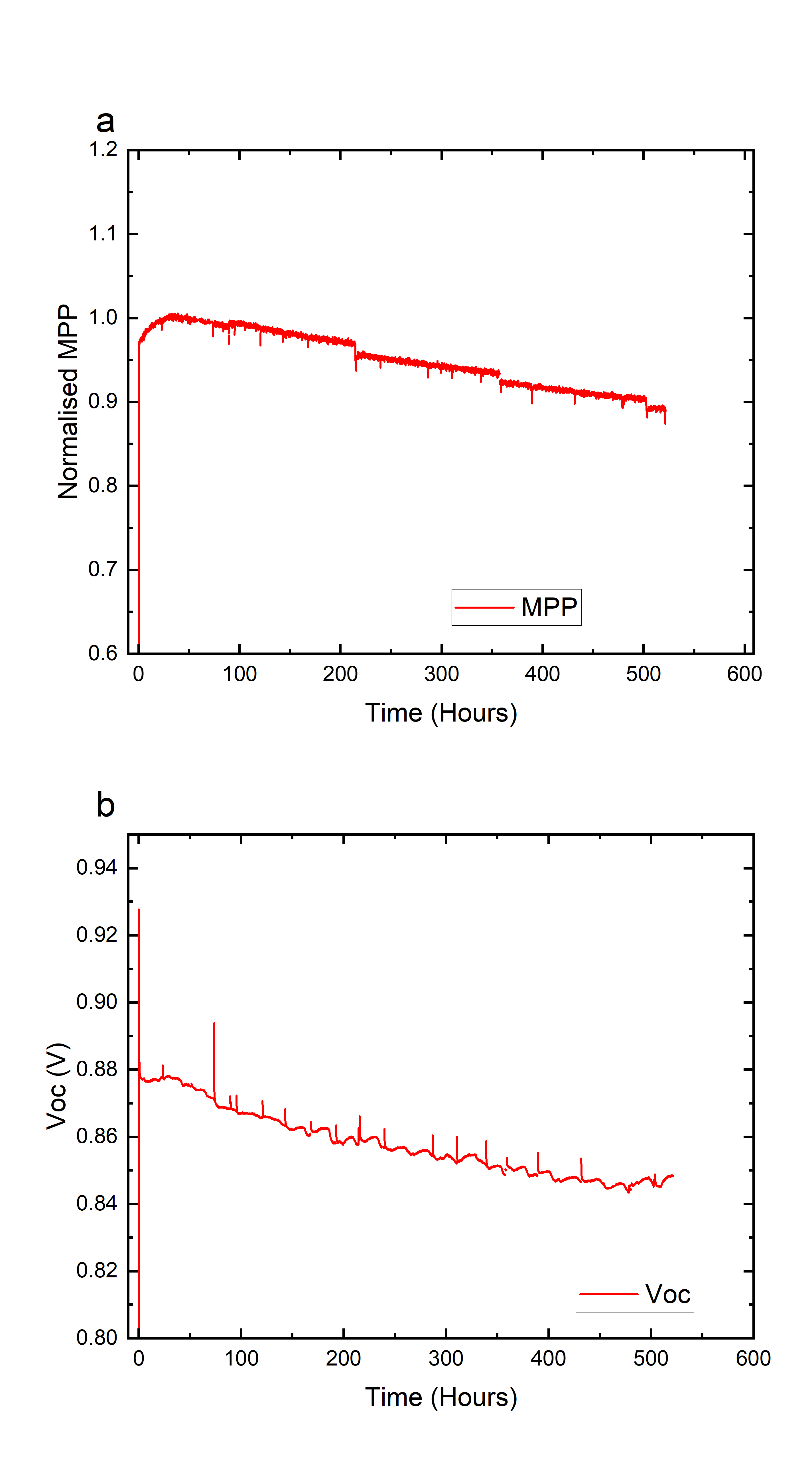}
    \caption{Aging data for devices SN112 and SN114, stressed under MPP and $\rm V_{OC}$, respectively. Both devices were aged for over 550 hours (~23 days).}
    \label{fig:aging_data}
\end{figure}

\begin{figure}[t]
    \centering
    \includegraphics[width=\linewidth]{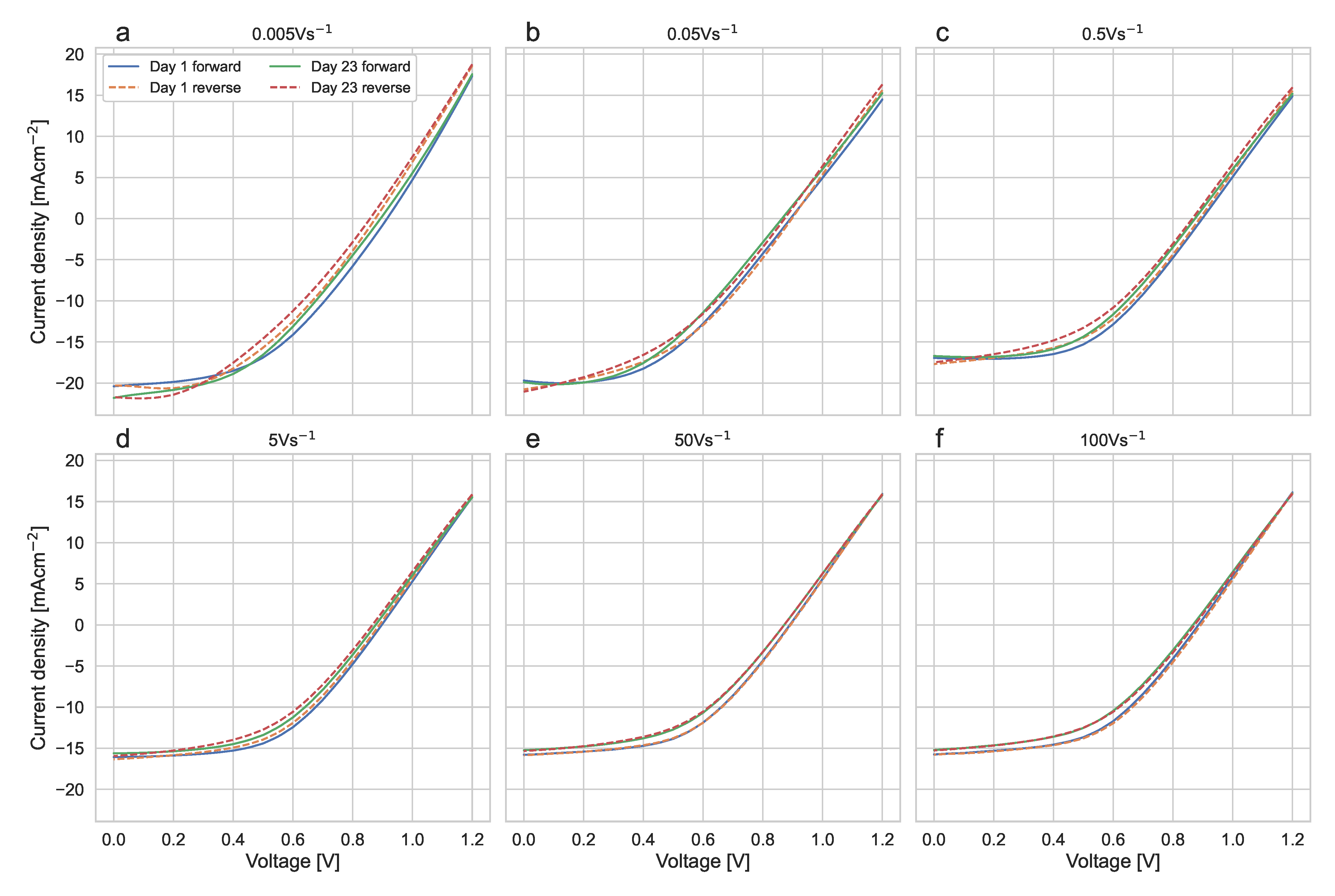}
    \caption{Comparison of the J-V curves by scan speed from the first and last day of the MPP-tracked device measured in the stability setup.}
    \label{fig:sn112jvhistfirstlast}
\end{figure}

\begin{figure}[t]
    \centering
    \includegraphics[width=\linewidth]{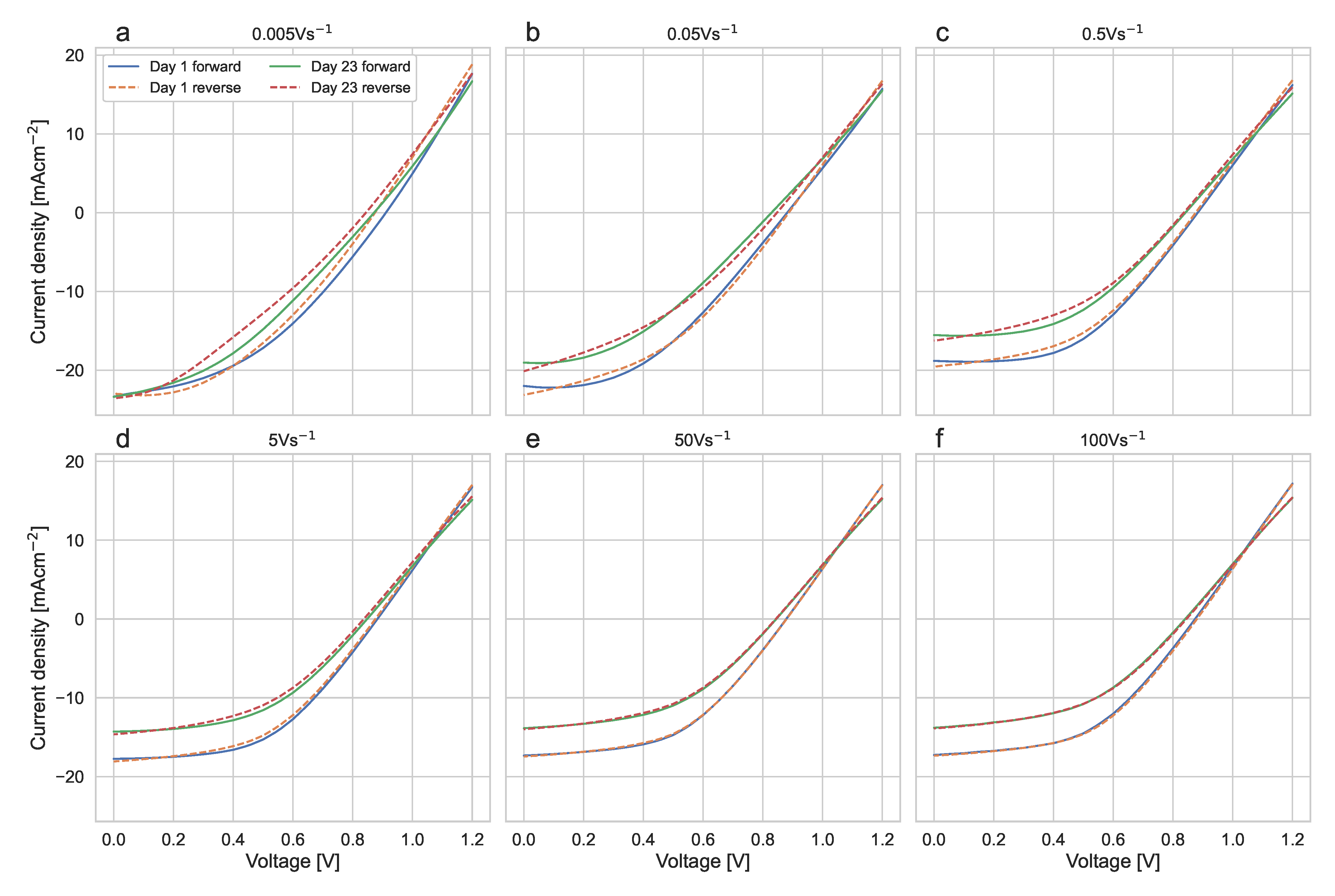}
    \caption{Comparison of the J-V curves by scan speed from the first and last day of the $\rm V_{OC}$-tracked device measured in the stability setup.}
    \label{fig:sn114jvhistfirstlast}
\end{figure}

The devices subjected to stress (SN112 at MPP and SN114 at $\rm V_{OC}$) exhibited clear signs of degradation. SN112 retained 90\% of its initial power conversion efficiency after aging (Figure \ref{fig:aging_data} a), while SN114, which was held at Voc, showed a reduction in open-circuit voltage from 0.88 V to 0.84 V. The devices were sufficiently aged to ensure that degradation processes occurred within their operational regime. As shown in Figure \ref{fig:aging_data} b, distinct spikes in Voc appear immediately after each J-V measurement. These transients are attributed to changes in the ionic distribution, consistent with observations reported in Reference \cite{https://doi.org/10.1002/aenm.202403850}, and the resulting ionic losses are also reflected in the J-V curves in Figures \ref{fig:sn112jvhistfirstlast} \& \ref{fig:sn114jvhistfirstlast}. The $\rm J_{SC}$ shows scan rate-dependent behaviour, where the losses are more prominent in the fastest scans.  This can also be seen as the spikes in Figure \ref{fig:aging_data} b.

\begin{figure*}[t]
    \centering
    \includegraphics[width=\linewidth]{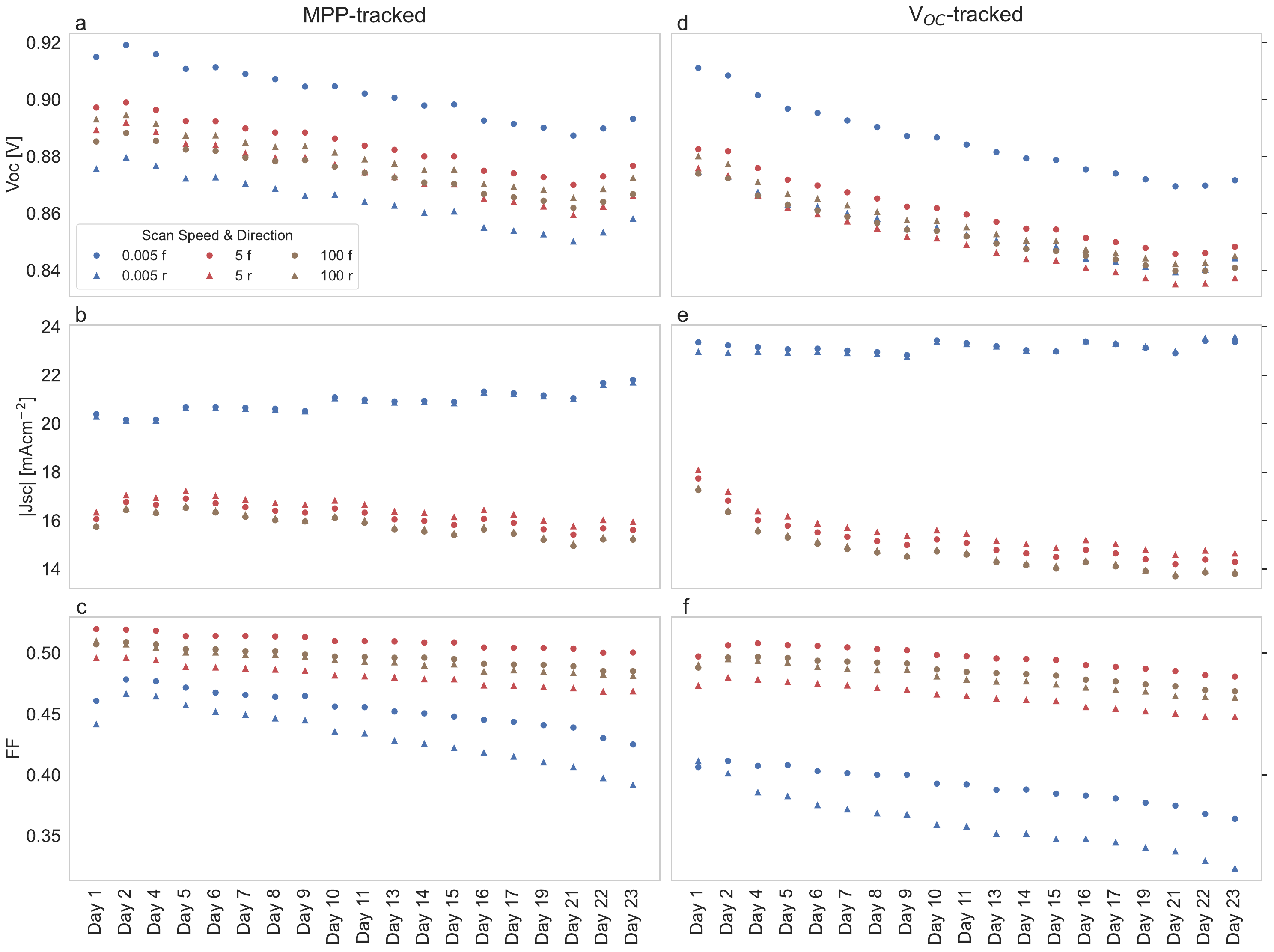}
    \caption{Evolution of key parameters for the two devices in the stability setup for a selection of the scan speeds. a-c show the results for the device kept at MPP, d-f for the device kept at $\rm V_{OC}$. a \& d: $\rm V_{OC}$. b \& e: Absolute value of $\rm J_{SC}$. c \& f: FF. Figure \ref{fig:allvocabsjscff} shows the full data for all scan speeds.}
    \label{fig:vocabsjscff}
\end{figure*}

\begin{figure*}[h]
    \centering
    \includegraphics[width=\linewidth]{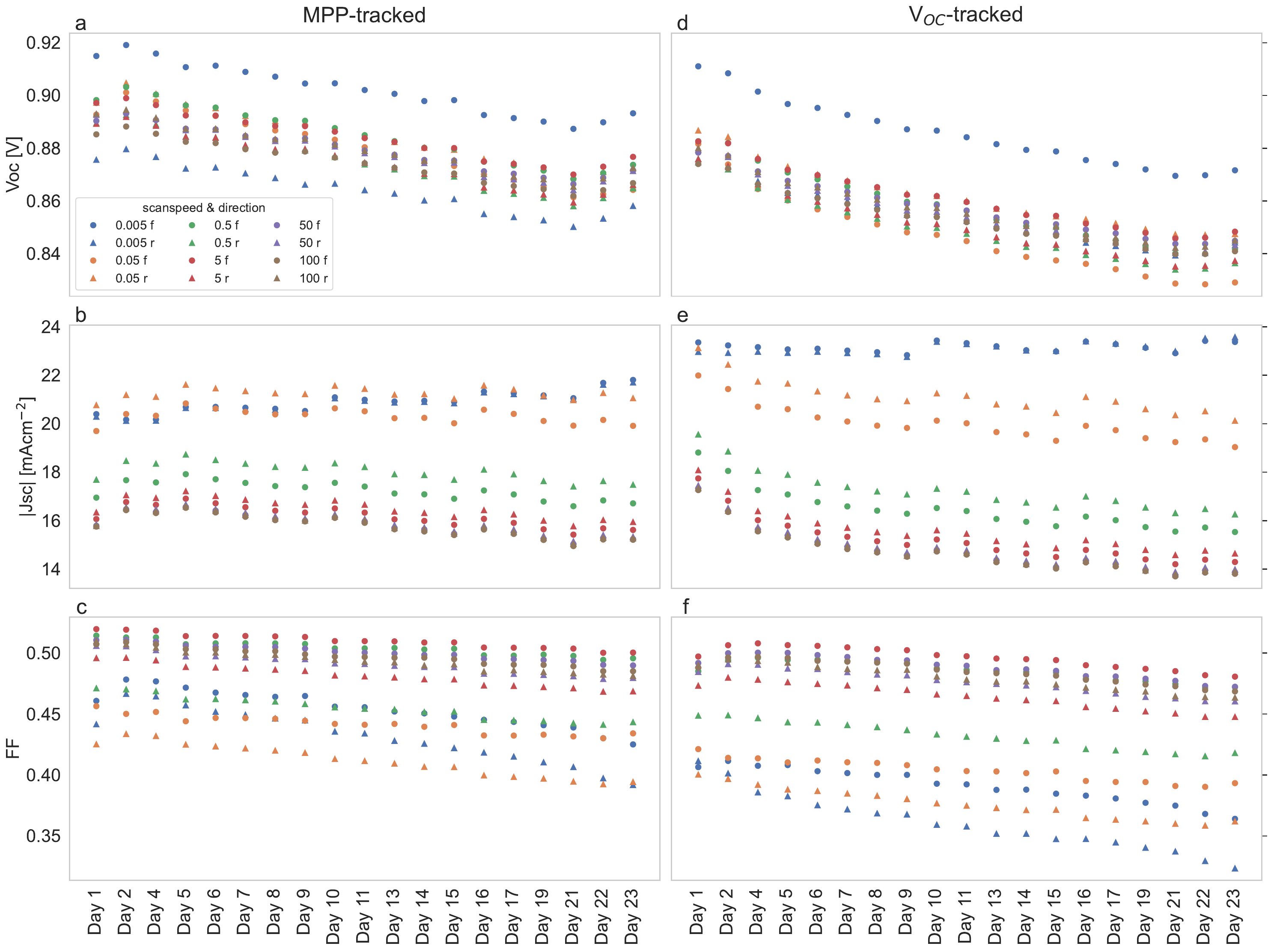}
    \caption{Complementary to Figure \ref{fig:vocabsjscff}. a-c show the results for the device kept at MPP, d-f for the device kept at $\rm V_{OC}$. a \& d: $\rm V_{OC}$. b \& e: Absolute value of $\rm J_{SC}$. c \& f: FF.}
    \label{fig:allvocabsjscff}
\end{figure*}

Before the parameter estimates from the encoder are discussed, the behaviour of $\rm V_{OC}$, $\rm J_{SC}$, FF, hysteresis and the series resistance of the measured J-V curves are analyzed and discussed.
Figures \ref{fig:vocabsjscff} \& \ref{fig:allvocabsjscff} a, d show that $\rm V_{OC}$ decreased over time for both devices similarly, but showing an increasing tendency for the last $\approx50$ hours. $\rm V_{OC}$ is scanning direction dependent in all cases, but less pronounced at higher scan speeds, as expected.
The behaviour of $\rm J_{SC}$ in the devices is not exactly the same, even though the end result is similar. It decreased in magnitude for both devices when the first and last days are compared for most scan speeds, but there are notable differences in between. The device that was kept at $\rm V_{OC}$ experienced a rather steep decrease of $\rm J_{SC}$ (Figures \ref{fig:vocabsjscff} \& \ref{fig:allvocabsjscff} b) at the beginning, for all scan speeds, except the slowest two ($0.05$ V/s \& $0.005$ V/s), with a trend that indicates some sort of saturation. The $0.05$ V/s scans still show the same trend, but always at higher currents. There is almost no change in $\rm J_{SC}$ for the $0.005$ V/s. Scanning direction dependencies of $\rm J_{SC}$ only seem to be present for ($0.05$, $0.5$, $5$) V/s, not for ($0.005$, $50$, $100$) V/s.
On the other hand, the device kept under MPP first experienced an increase in $\rm J_{SC}$ before they decreased as well (Figures \ref{fig:vocabsjscff} \& \ref{fig:allvocabsjscff} b), except for the $0.005$ V/s and $0.05$ V/s cases. The slowest scan speed shows an improvement of $\rm J_{SC}$ over time without scan speed dependence, and the $0.05$ V/s measurements are relatively stable. %\textit{Interestingly, the measurements with the slowest scan speeds show much higher currents than the rest in both devices \textcolor{red}{why?, because of ionic distribution??}, and they show a more stable behaviour. \textcolor{red}{still needed? Probably not!}}
The FFs show similar trends of a slight decrease over time for both devices (Figures \ref{fig:vocabsjscff} \& \ref{fig:allvocabsjscff} c, f). The difference between forward and reverse scan is almost negligible for the fastest scan speeds, and very prominent for the slowest ones, which is explained by the differences in $\rm V_{OC}$. The picture for $0.5$ V/s is somewhat inconclusive for both devices, because the forward scan follows the trend of the faster scans with only slightly decreased FF.% but the reverse scan is closer to the slower scan speed trends.

\begin{figure}[t]
    \centering
    \includegraphics[width=\linewidth]{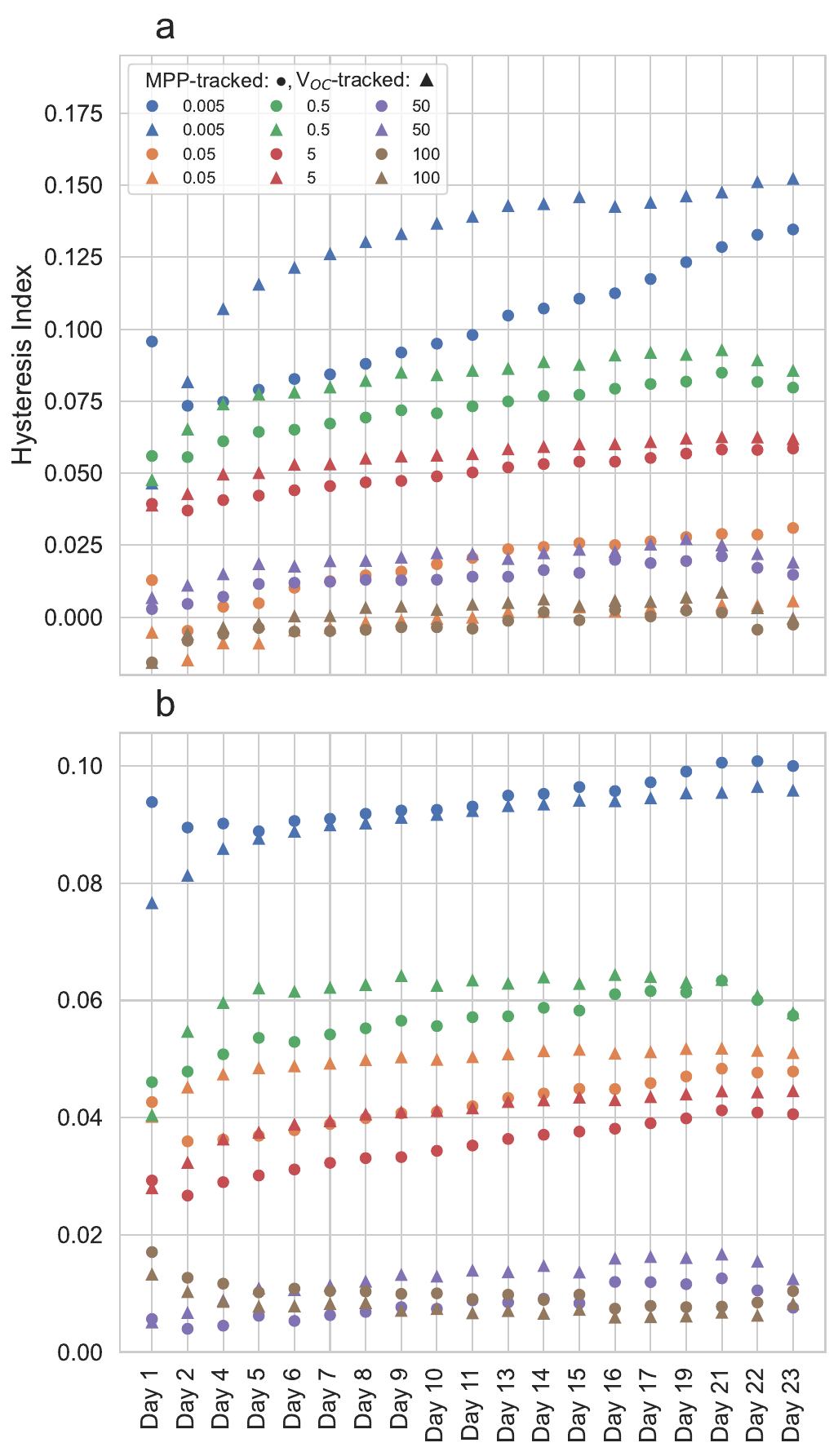}
    \caption{Two different ways to determine hysteresis. a: Hysteresis (PCE). b: Ratio of difference in area of J-V curves at each scan speed.}
    \label{fig:bothystereses}
\end{figure}

%Regarding hysteresis, the scan speeds ($0.05$, $50$, $100$) V/s show almost no hysteresis, some of them start with normal (reverse-forward) hysteresis, but show negative hysteresis with time. For the fastest scan speeds, this indicates that they are already too fast to resolve ionic effects. In case of $0.05$ V/s, this is more difficult to interpret since the lower scan speed ($0.005$ V/s) scan speed shows the strongest hysteresis. Forward and reverse J-V curves intersect multiple times at scan speeds of $0.05$ V/s, which cannot be resolved by the hysteresis index. For the other sacan speeds, hysteresis is always negative, decreasing with time as well. Both devices show very similar trends, as can be seen in Figure \ref{fig:hysteresis}. The only outlier is the slowest scan speed ($0.005$ V/s), where the difference is somewhat larger.
Due to ionic charge redistribution, hysteresis can be observed in PSCs \cite{hyster_tress_1,richardson_hyster,tress_metal_2017}, which can also be seen in Figures \ref{fig:sn112jvhistfirstlast} \& \ref{fig:sn114jvhistfirstlast} for the scan speeds ($0.005-5$) V/s. This hysteresis is oftentimes reported with a dimensionless hysteresis index that quantifies the difference between forward and reverse scan, for example by $\rm\frac{reverse\, PCE-forward\, PCE}{reverse\, PCE}$. A positive index, with higher currents in reverse direction, is normal, a negative index inverted hysteresis. This definition can be problematic when one wants to quantify the behaviour of J-V curves that show both, normal and inverted hysteresis at different voltages, because the effect partially cancels out. This can be observed in both devices at the scan speed $0.05$ V/s (Panel b in Figures \ref{fig:sn112jvhistfirstlast} \& \ref{fig:sn114jvhistfirstlast}).
Figure \ref{fig:bothystereses} therefore shows two different hysteresis indices, panel a is the hysteresis based on PCE, panel b the hysteresis from the area difference of the J-V curves. Both representations have advantages and disadvantages. Panel a contains information about normal or inverted hysteresis, but only for one point, and panel b shows the absolute area between the forward and reverse scan, but does not reveal if hysteresis is inverted or not.
The scan speeds ($0.05$, $50$, $100$) V/s show almost no hysteresis in panel a. This is consistent with the results in panel b for the fast scan speeds, but as already indicated, not for $0.05$ V/s.

%Some of the J-V curves start with normal hysteresis $\rm\frac{reverse-forward}{reverse}>0$, but then change to inverted hysteresis with time.
In the cases of the fast scan speeds, this indicates that they are already too fast to resolve ionic effects, but for $0.05$ V/s, the result is inconclusive
For the other scan speeds, hysteresis is always inverted, increasing in magnitude over time. Both devices show very similar trends. The only outlier is the slowest scan speed, where the difference between MPP and $\rm V_{OC}$ tracked devices is more notable. However, the changes are still not dramatic as can be seen in Figures \ref{fig:sn112jvhistfirstlast}, \ref{fig:sn114jvhistfirstlast}.
%\textcolor{red}{for hysteresis interpretation probably read https://www.nature.com/articles/s41467-024-53162-z but the jv curves there do not really match our measurements}

Summarizing the discussed findings, it can be said that when stressing these devices, in both cases, $\rm V_{OC}$ changes notably. $\rm J_{SC}$ also decreases substantially for most scan speeds when the device is kept at $\rm V_{OC}$, but it is relatively stable under MPP tracking. FF tends to decrease with time, but not substantially.
%The question is whether one can already identify possible signs of degradation or at least change over time on the basis of these findings. For this, some degree of expert knowledge is necessary.
Different scan speeds show different trends, as mentioned before, which is not easy to quantify consistently when we have multiple scan speeds. For single J-V curves, like steady state, a discussion of the influence of different parameters can be found in literature \cite{meins_1, thesejv} (in the respective SI). Of course, not every possible parameter is varied there, so it may be possible that other changes could lead to such a change in the J-V curve as well. Also note that there are differences between the carbon-based device here to the planar architecture simulated in \cite{meins_1, thesejv}, but the influence of parameter changes is basically the same. Out of all the parameters discussed, there are only a few that show the behaviour observed in the studied devices. These parameters are predominantly related to recombination, in the bulk or at the surface. In principle, these changes can, to some extent, be explained by changes in ion densities as well. Since the influence on the J-V curve can depend on the ion species \cite{thesejv}, which was not included in the simulation originally used to train the AE ($\rho_A=\rho_C$), this is not covered by the AE model. Based on this information, it is expected that, if the AE estimates the parameters reasonably, the most notable changes should be related to recombination.

\begin{figure*}[t]
    \centering
    \includegraphics[width=1\linewidth]{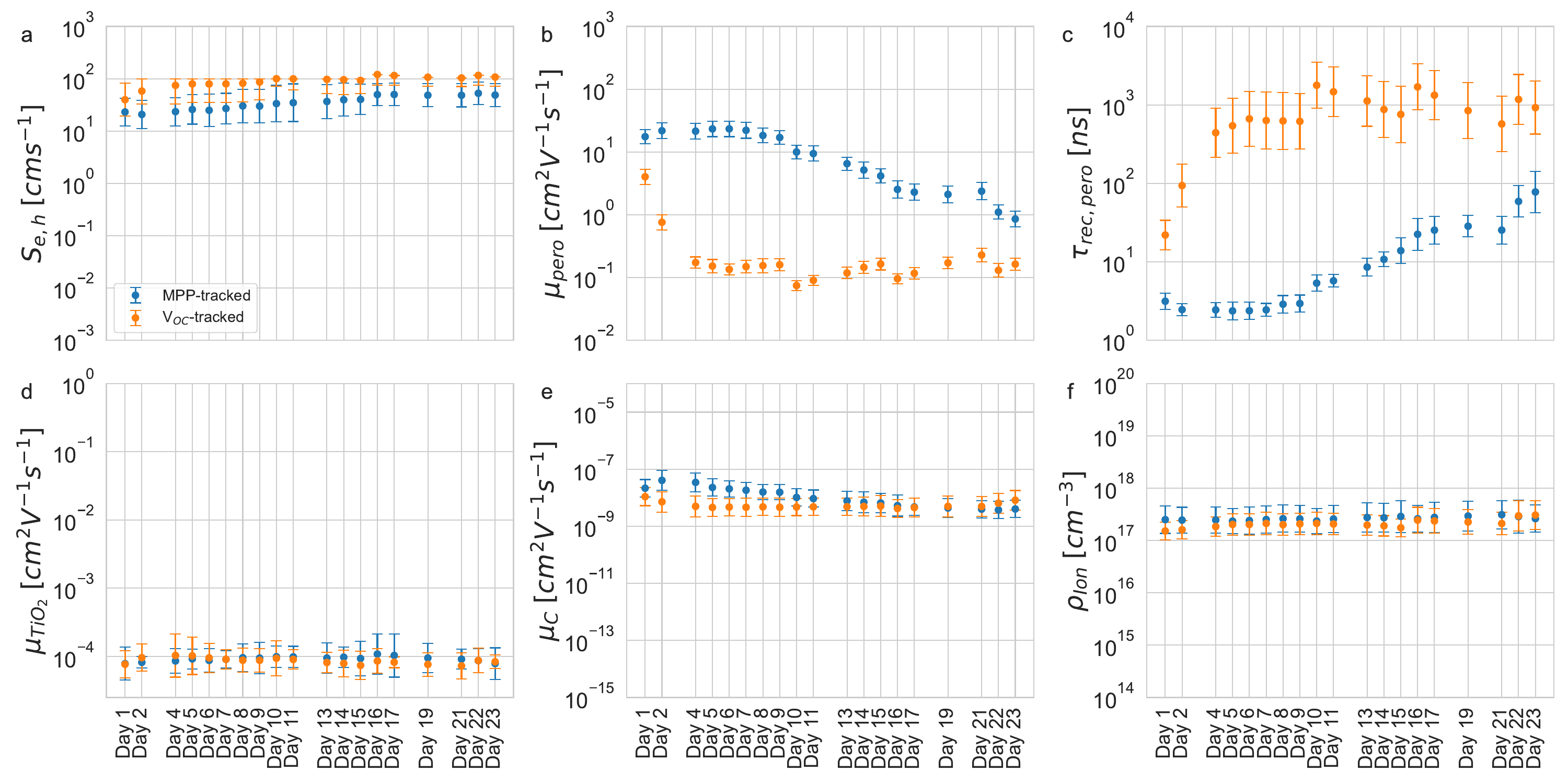}
    \caption{Evolution of AE-estimated parameters, based on the J-V curves measured on each day. Uncertainties are estimated based on the mean difference between true parameter values of the test set in Figure \ref{fig:transtestparam} and their estimates in an interval of $\pm0.02$ for each predicted device parameter estimate.}
    \label{fig:param_est_secondstudy}
\end{figure*}

The AE-based parameter estimates inferred from the  J-V curves measured while the devices were stressed are shown in Figure \ref{fig:param_est_secondstudy} and Table \ref{tab:encodedparams}. These results are now interpreted and, where possible, supported with different characterization measurements before and after stressing the devices. 

\begin{figure}[t]
    \centering
    \includegraphics[width=1\linewidth]{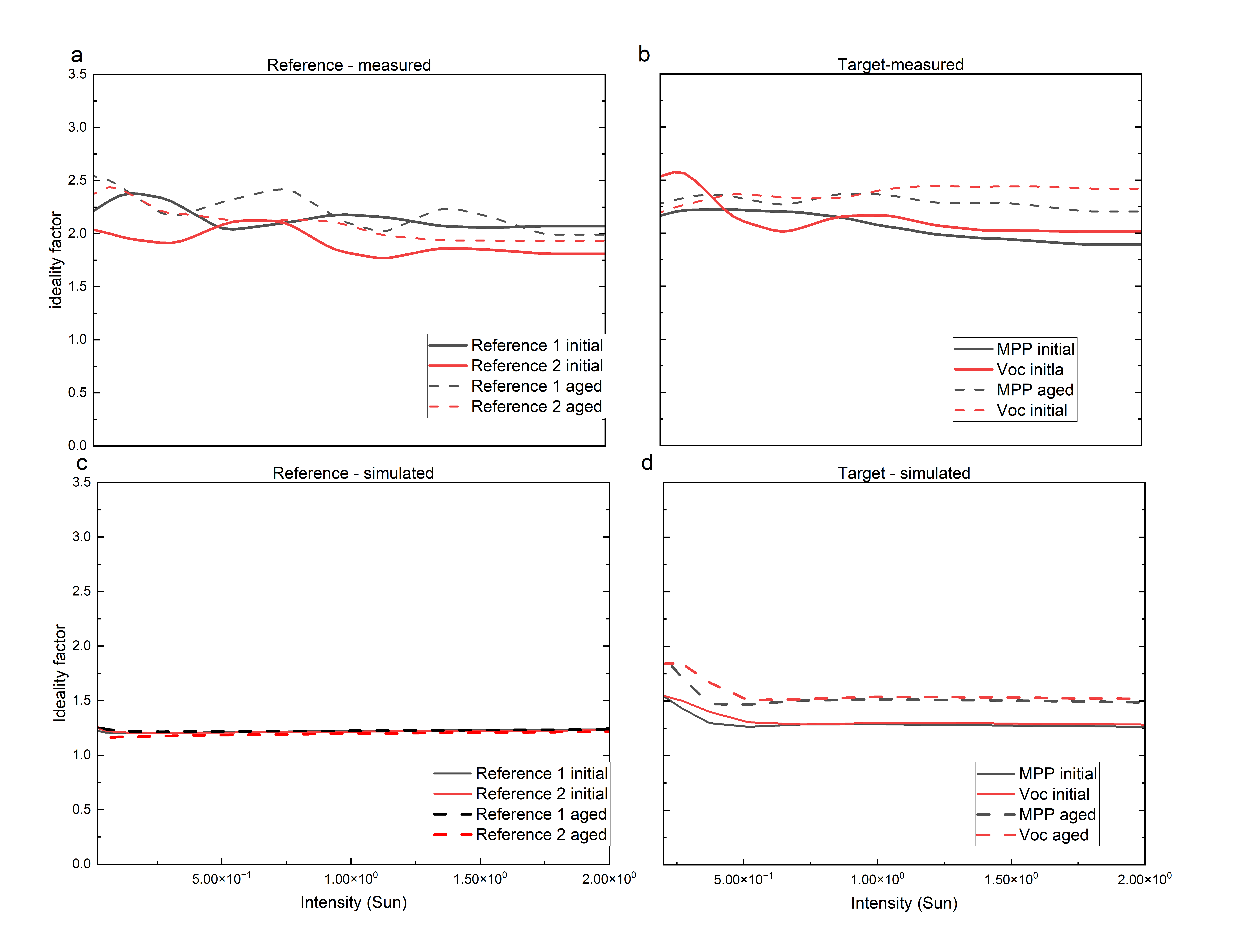}
    \caption{ a,b shows the measured ideality factors of reference and stressed devices. c, d are the simulated ideality factors of the reference and stressed devices}
    \label{fig:ideality}
\end{figure}

\textbf{$S$}: The surface recombination velocity increases over time for both devices, and it seems that it approaches a value of around $50$ cm/s for the device kept at MPP and around $100$ cm/s for the device kept at $\rm V_{OC}$. The initial difference of the parameter estimates for the two devices shows some amount of sample variation, which can be seen in the J-V curves as well. The changes over time are not the same, which could be related to the different stressing conditions. Light-intensity-dependent J–V measurements were performed to extract the experimental ideality factor, which was further compared with the values obtained from drift–diffusion simulations using parameters estimated by the autoencoder model as depicted in Figure \ref{fig:ideality}. While the simulated ideality factors, derived from AE-parameterised drift–diffusion modelling, reproduce the qualitative trend of the measurements, their absolute values are smaller, which may arise from the neglect of ionic redistribution and associated field screening effects that enhance recombination in real devices \cite{PhysRevApplied.11.044005}. %As a consequence, the effect of a significant increase of $S$ follows.

\begin{figure*}[t]
    \centering
    \includegraphics[width=\linewidth]{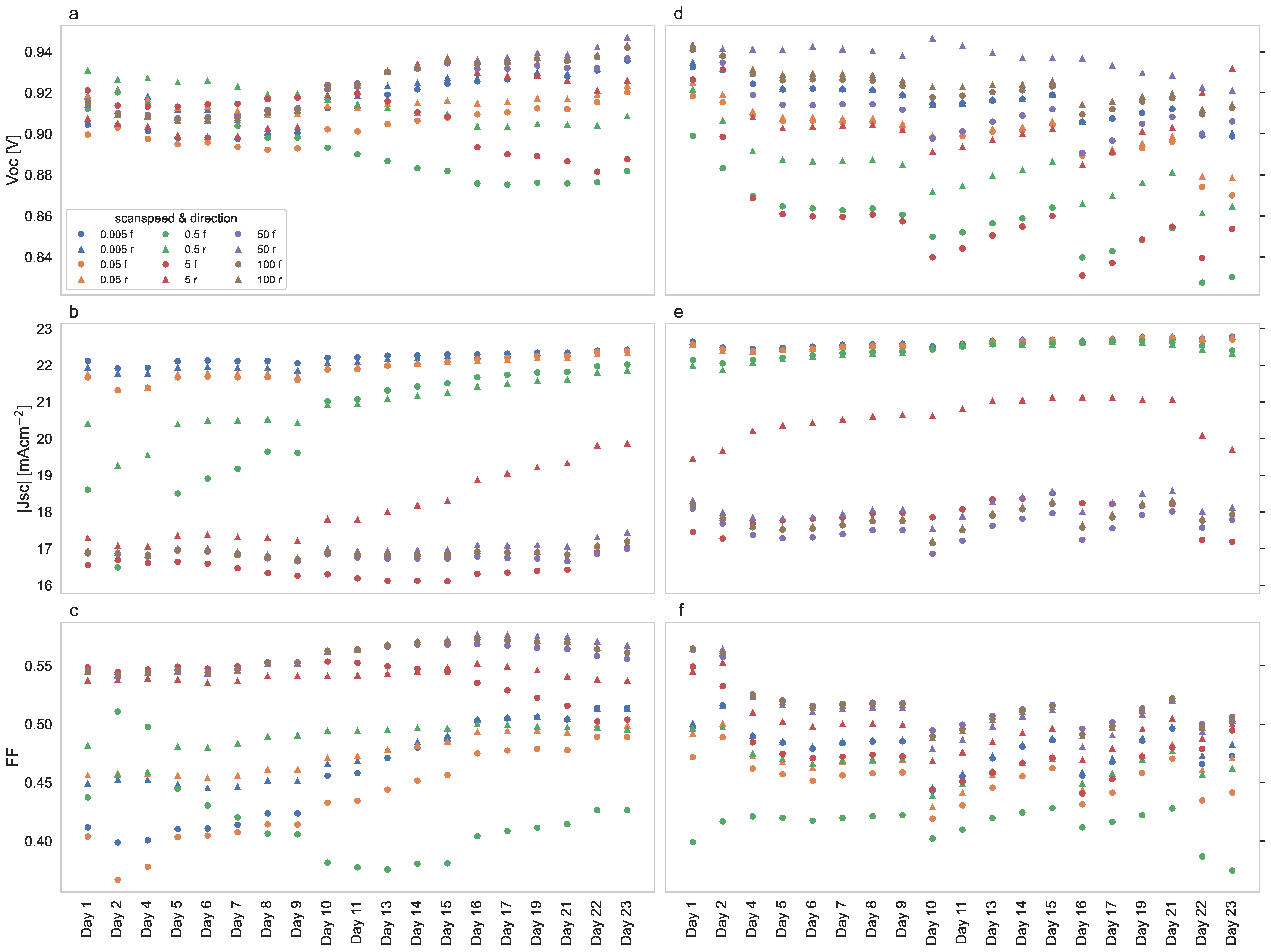}
    \caption{Evolution of key parameters extracted from simulations based on parameter estimates. a-c show the results for the device kept at MPP, d-f for the device kept at $\rm V_{OC}$. a \& d: $\rm V_{OC}$. b \& e: Absolute value of $\rm J_{SC}$. c \& f: FF.}
    \label{fig:vocjscffsim}
\end{figure*}

\textbf{$\mu_{pero}$}: As no significant decrease in $\mu_{TiO_2}$ is observed, while the series resistance shows a consistent rise over time, it is reasonable to attribute the emerging transport limitations to changes within the perovskite layer. Although the $\rm V_{OC}$ of these devices does not appear to be bulk-limited, the evolution of the series resistance aligns closely with $\mu$ of the device aged under MPP conditions. The device aged under Voc exhibits a similar trend up to day 9, after which the alignment diverges (see Figure \ref{fig:vocjscffsim}). These effects are likely not related to $\mu_{pero}$, but will be disscussed in the section about ions. Notably, by this stage, the extracted $S$ parameter exceeds the range represented in the AE model’s training dataset, indicating that the system has entered a regime not captured during model development.

\begin{figure*}[t]
    \centering
    \includegraphics[width=1\linewidth]{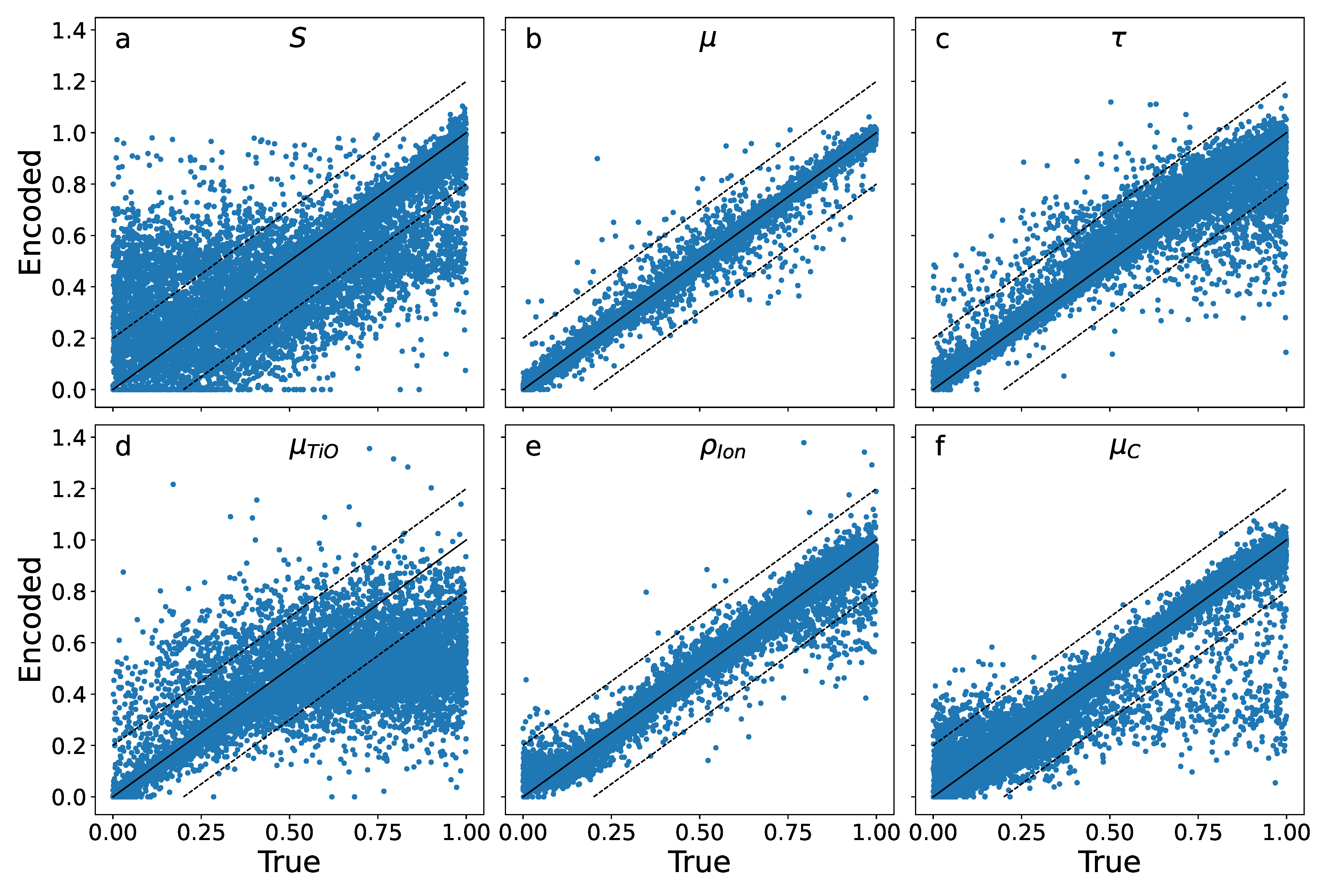}
    \caption{Log-normalized parameter estimates for the test set. a: $S$, b: $\mu$, c: $\tau$, d: $\mu_{TiO_2}$, e: $\rho_{Ion}$, f: $\mu_C$. The dashed lines indicate the boundaries of a confidence interval where the residuals are $\leq 0.2$.}
    \label{fig:transtestparam}
\end{figure*}

\textbf{$\tau$}: For the device kept at MPP, $\tau$ is, initially, within the same order of magnitude with what was obtained from TrPL, which was fitted assuming a double exponential decay. Both devices had decay time constants of $1.47$ ns and $1.39$ ns, respectively. These values are in agreement with what was measured elsewhere \cite{carbonrecord}. However,  inferring $\tau$ from transient photovoltage (TPV) measurement, the results are several orders of magnitude higher, which is oftentimes the case in PSCs \cite{trpltpctpvtau,carbonrecord}. These measured effective lifetimes account for various effects, including bulk and surface recombination. The capacitive and ionic properties of the device dominate the measurements of TPV \cite{lifetimestuff,bisquert2021frequency}. The initially estimated low lifetimes can be explained by the mesoporous structures infiltrated with perovskite, leading to a bulk that is effectively one big interface with presumably high defect densities at these surfaces.
For the device kept at $\rm V_{OC}$, $\tau$ is already higher from the beginning, and increases to values that are more comparable to reported lifetimes based on TPV measurements. From Figure \ref{fig:transtestparam}, it can be seen that the AE predictions are more reliable at shorter lifetimes, because if $\tau$ is longer than the (extreme) time needed for extraction, it does not affect the J-V curve anymore, which is indicated by larger error bars in Figure \ref{fig:param_est_secondstudy}. Due to the significant differences between lifetimes depending on the measurements they are based on, it is difficult to tell which method is closer to reality. It is therefore challenging to tell if the predictions are correct.

\begin{figure}[t]
    \centering
    \includegraphics[width=\linewidth]{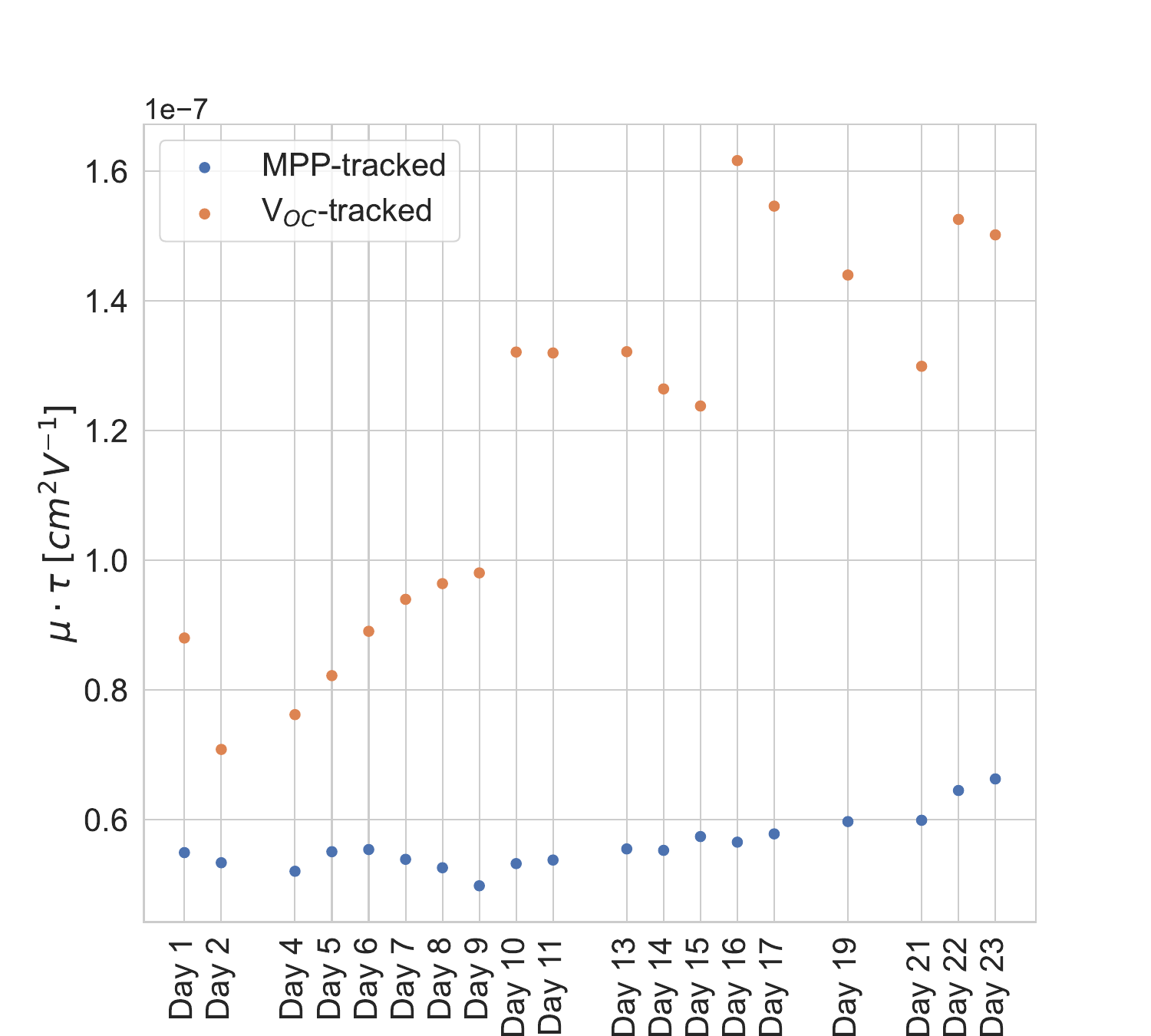}
    \caption{The evolution of $\mu\tau$ is shown in this figure, complementary to the discussion and Figure \ref{fig:param_est_secondstudy}.}
    \label{fig:mutau}
\end{figure}

\textbf{$\mu_{pero}$} \& \textbf{$\tau_{\rm rec,\,pero}$}: These two parameters are also discussed together because the product is closely related to the diffusion length $L=\sqrt{D\tau}=\sqrt{\mu\tau\frac{k_BT}{q}}$. The product $\mu\tau$ mainly influences $\rm J_{SC}$, and, if $S$ dominates, the effects of $\tau$ are suppressed. For the device kept at MPP, the AE estimates show a decrease of $\mu_{pero}$ and an increase of $\tau_{\rm rec,\,pero}$ of around the same order of magnitude, leading to a constant product $\mu\cdot\tau$ (Figure \ref{fig:mutau}), closely following $\rm J_{SC}$ of the slowest scan speed in Figure \ref{fig:vocabsjscff}b. The results for the $\rm V_{OC}$-tracked device are not as conclusive on first sight because the measured $\rm J_{SC}$ (Figures \ref{fig:vocabsjscff} \& \ref{fig:allvocabsjscff} e) is constant over time and does not follow the behaviour shown if Figure \ref{fig:mutau}.
%On the other hand, $S$ increases much more in case of the $\rm V_{OC}$-tracked device, acting against an improved current. The effective charge carrier lifetime is reduced \cite{taueff_1,taueff_2,taueff_3},

%\begin{equation}
%    \frac{1}{\tau_{eff}} = \frac{1}{\tau_{rec,\,pero}} + \frac{2S}{d},
%\end{equation}

%where $d$ is the absorber thickness. This $\tau_{eff}$ balances the positive effect of an enhanced $\mu\tau$, leading to a constant $\rm J_{SC}$.
%%On the other hand, in case of $\rm V_{OC}$ tracking, there is a significant change, it is estimated that $\mu\cdot\tau$ almost doubles. Not continuously, but in several notable jumps between two consecutive days. \textcolor{red}{where can this behaviour be seen?}

\textbf{$\mu_{TiO_2}$}: These parameter estimates are very similar for both devices and constant over time, changes only occur within the calculated uncertainties. It seems that $\mu_{TiO_2}$ is not affected by this type of stress. Such behaviour is consistent with the relatively stable nature of the compact $\mu_{TiO_2}$ layer, where electron transport is governed by trap-limited diffusion \cite{CHE20171064}. Note that only slight changes of $\mu_{TiO_2}$ can already influence the J-V curve substantially. Consequently, simulation based on the estimated parameters can agree much better with the measurement if $\mu_{TiO_2}$ is varied only within the uncertainty range.

\begin{figure}[t]
    \centering
    \includegraphics[width=1\linewidth]{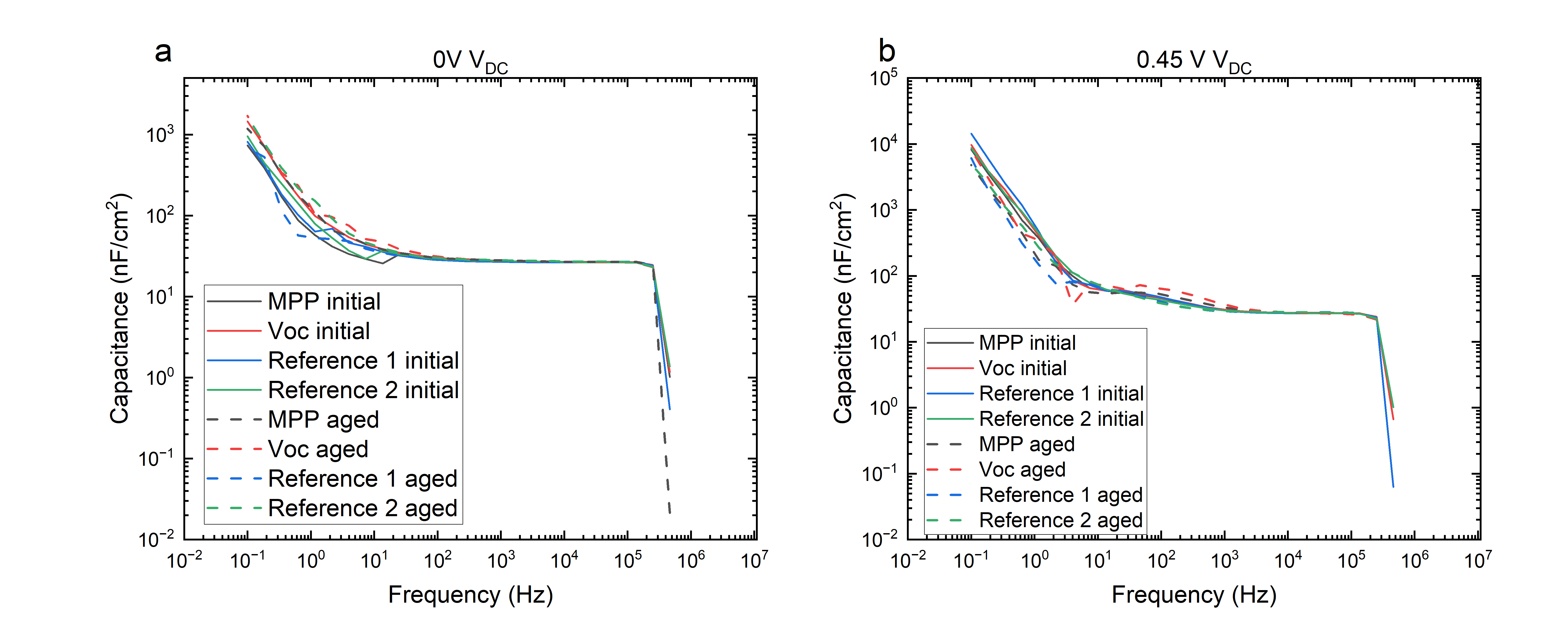}
    \caption{Capacitance vs frequency plot of cell in dark before and after aging, along with references. Electrochemical impedance spectroscopy (EIS) was conducted in the dark using the PAIOS system. A 10 mV AC perturbation was applied across a frequency range of 10 MHz to 0.1 Hz to characterize the frequency-dependent electrical response of the devices.}
    \label{fig:EIS}
\end{figure}

\textbf{$\mu_C$} \& \textbf{$\rho_{A,\,C}$}: Overall, it seems that there is not much change in the ionic parameters. It seems that for the device kept at MPP, $\mu_C$ decreases slightly. The changes are within the indicated uncertainties, or just slightly outside. For the device kept at $\rm V_{OC}$, the variation in $\mu_C$ is within uncertainty over all $23$ days. For both devices, changes of $\rho_{A,\,C}$ are within uncertainty, indicating approximately constant ion densities. However, it seems that the operational bias at $\rm V_{OC}$ accelerates the series resistance increase, likely due to accumulation of ions at the interface, leading to FF loss \cite{thiesbrummel_ion_induced_2024}. The behaviour of these parameters during the experiment, especially $\mu_C$, is supported by the hysteresis that is not much affected (Figure \ref{fig:bothystereses}), and the small estimated changes in the J-V measurements (Figures \ref{fig:sn112jvhistfirstlast}, \ref{fig:sn114jvhistfirstlast}) could already indicate that $\mu_C$ is indeed not exactly constant over time. Comparing the J-V curves from the first and last day of the study (Figures \ref{fig:sn112jvhistfirstlast}, \ref{fig:sn114jvhistfirstlast}), it can be seen that the shapes of the J-V curves are not much affected.
A quantitative verification of the ionic density is not possible from impedance spectroscopy, since the plateau at lower frequencies in the capacitance-frequency (C-$f$) plot (Figure \ref{fig:EIS}) that corresponds to the ionic capacitance\cite{plateau} could not have been reached. But there is no significant change in the C-$f$ plot before and after stressing, which points to minimal variation in ion density. This agrees with the ionic density estimates we get from the AE.

\begin{table*}[t]
    \centering
    \caption{Parameter estimations from the Paios measurements before and after aging.}
    \label{tab:paiosparams}
        \begin{adjustbox}{width=1\textwidth}
        \begin{small}
        \begin{sc}
        \begin{tabular}[htbp]{llrrrrrr}

        \toprule
        & & $\rm S$ $\rm [cms^{-1}]$ & $\rm \mu$ $\rm [cm^2V^{-1}s^{-1}]$ & $\tau$ $\rm [ns]$ & $\mu_{TiO_2}$ $\rm [cm^2V^{-1}s^{-1}]$ & $\rm\mu_C$ $\rm [cm^2V^{-1}s^{-1}]$ & $\rm\rho_{Ion}$ $\rm [cm^{-3}]$ \\
         
        \midrule

        Reference 1 & before & $2.84$ & $13.32$ & $4.16$ & $3.57\cdot10^{-5}$ & $1.12\cdot10^{-7}$ & $1.82\cdot10^{17}$ \\

        & after & $9.85\cdot10^{-3}$ & $19.26$ & $10.09$ & $7.89\cdot10^{-5}$ & $2.12\cdot10^{-5}$ & $1.11\cdot10^{17}$ \\

        \midrule
        
        Reference 2 & before & $5.33\cdot10^{-1}$ & $12.88$ & $3.87$ & $3.59\cdot10^{-5}$ & $9.30\cdot10^{-8}$ & $1.79\cdot10^{17}$ \\

        & after & $1.21\cdot10^{-3}$ & $15.33$ & $8.93$ & $8.06\cdot10^{-5}$ & $4.60\cdot10^{-5}$ & $2.47\cdot10^{17}$ \\
        
        \midrule
        
        MPP-tracked & before & $5.98$ & $16.21$ & $6.71$ & $3.64\cdot10^{-5}$ & $1.10\cdot10^{-7}$ & $1.00\cdot10^{17}$ \\

        & after & $52.03$ & $6.47\cdot10^{-1}$ & $1.74\cdot10^2$ & $5.85\cdot10^{-5}$ & $2.03\cdot10^{-8}$ & $1.15\cdot10^{17}$ \\

        \midrule

        $\rm V_{OC}$-tracked & before & $12.79$ & $11.93$ & $4.30$ & $3.85\cdot10^{-5}$ & $6.25\cdot10^{-8}$ & $1.17\cdot10^{17}$ \\

        & after & $67.86$ & $4.21\cdot10^{-1}$ & $2.16\cdot10^2$ & $5.62\cdot10^{-5}$ & $6.28\cdot10^{-9}$ & $1.32\cdot10^{17}$ \\

        \bottomrule
    \end{tabular}
    \end{sc}
    \end{small}
    \end{adjustbox}
    \vskip -0.1in
\end{table*}

The estimated parameters can be used as simulation input to compare measurements and simulations. For the device kept at MPP, the model agrees initially, slightly overpredicting the current and FF. These differences are approximately constant for the first $9$-$11$ days, depending on the scan speed. As time passes, the differences between the measurements and simulations increase, where the scan speeds ($0.005$, $0.05$, $50$, and $100$) V/s do not show hysteresis in simulation, and the deviations are consistent for these scan speeds. For $0.5$ V/s and $5$ V/s, the results are a bit different. For $0.5$ V/s, the current matches the experiment in forward much better between MPP and $\rm V_{OC}$, but overpredicts the current between MPP  and $\rm J_{SC}$. At $5$ V/s, the simulation matches the measurement almost perfectly throughout the 23 days. On the other hand, with increasing time, the difference between the measurement and simulation of the reverse scanning direction increases for these two scan speeds, current and FF are overpredicted, as in the other cases. With higher scan speeds, $\rm V_{OC}$ is also estimated to be too high. The trends in the simulations for the device kept at $\rm V_{OC}$ are very similar, but they are more pronounced. These differences indicate that the estimated values or the underlying DD model are not as close to the real devices as intended, and there are several possible reasons for this. Consulting the literature again to discuss the J-V curves \cite{meins_1, thesejv}, these differences can be explained by changes in ion densities or parameters related to recombination in the bulk or at the surface. These two different possibilities are discussed separately. It has recently been found that ionic effects are the dominating factor of light-induced degradation of PSCs \cite{thiesbrummel_ion_induced_2024}. There, it has been shown that recombination in the bulk and at the surface depends on ion density, with a stronger influence on the surface recombination. The result, lower current and FF, is one mechanism that can change the simulated J-V curves towards what was measured. But why do the results not match? The AE could estimate higher ion densities and possibly solve the problem, but it does not. Why? Cations accumulate near the carbon-perovskite interface, causing recombination. The behaviour of this effect in a planar device, as simulated, is very different compared to the actual mesoporous structure. Furthermore, in a real device, the mobile ions move through the different layers filled with perovskite, in this case the two mesoporous TiO$_2$ and ZrO$_2$, as well as partially the carbon layer. The mesoporous TiO$_2$-perovskite layer is, effectively, one big surface, which cannot be simulated in a one-dimensional DD model for obvious reasons.

\section{Discussion}

In this study, it has been demonstrated that AEs can be used to track device changes during degradation on a fundamental level.
Two carbon-based PSCs were stressed for 23 days under light, one was kept at MPP, the other at $\rm V_{OC}$ in order to provoke degradation. While the device kept at MPP only changes slightly, the device kept at $\rm V_{OC}$ shows much stronger signs of degradation, especially with regard to $\rm J_{SC}$. 
Based on the evolution of $\rm V_{OC}$, $\rm J_{SC}$, and FF, it was discussed what quantities most likely changed based on classical J-V curve analysis.
The AE was trained to predict different material parameters from J-V curves as an input, and, based on the J-V curves measured during aging the devices under examination, estimated and evolution of physical parameters which is in agreement with what was expected classically. The results are, where possible, supported by additional measurements.

\begin{figure}[t]
    \centering
    \includegraphics[width=1\linewidth]{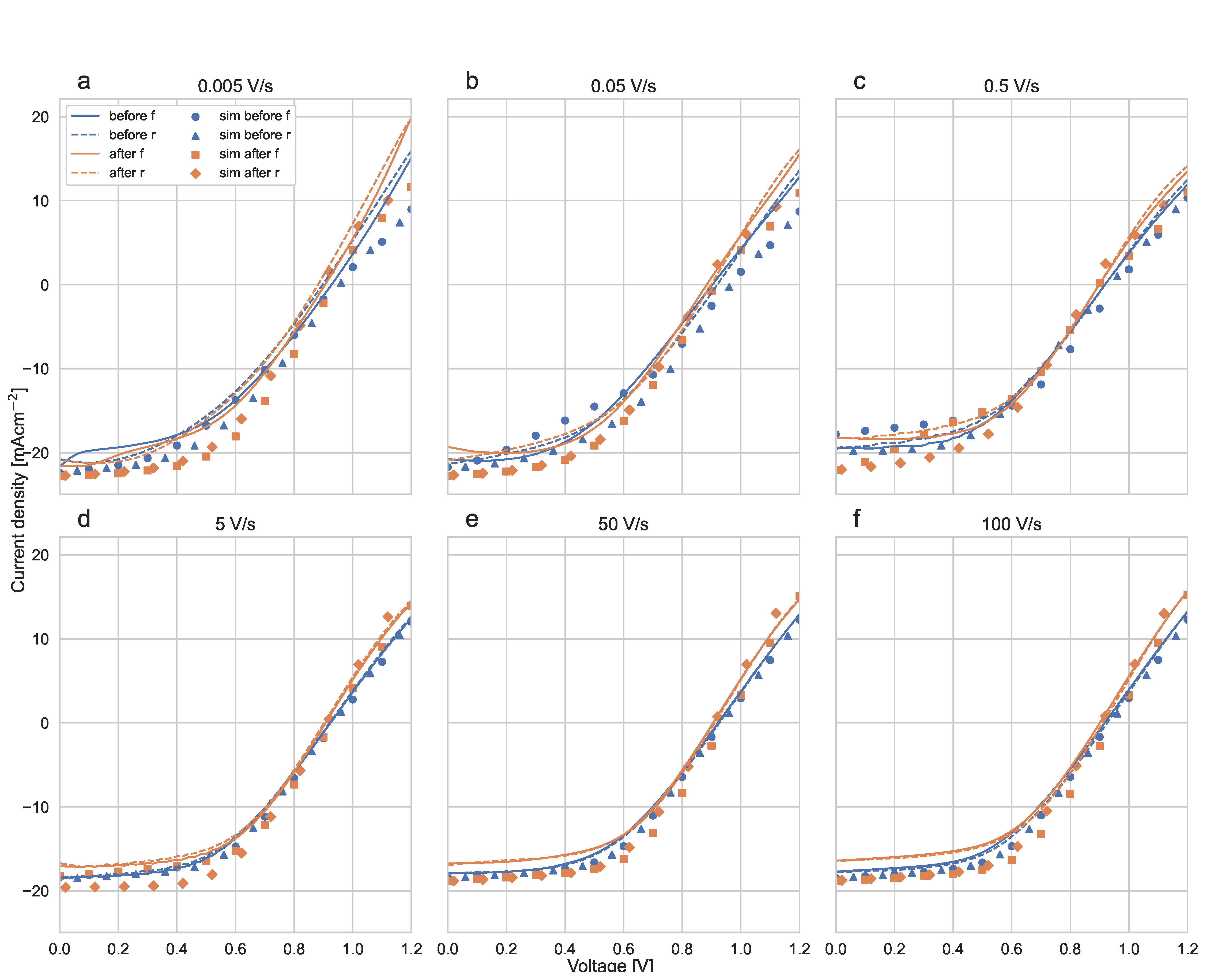}
    \caption{Measured J-V curves with Paios and simulation based on AE estimated parameters for the device kept at MPP before and after stressing.}
    \label{fig:paios_112}
\end{figure}

\begin{figure}[t]
    \centering
    \includegraphics[width=1\linewidth]{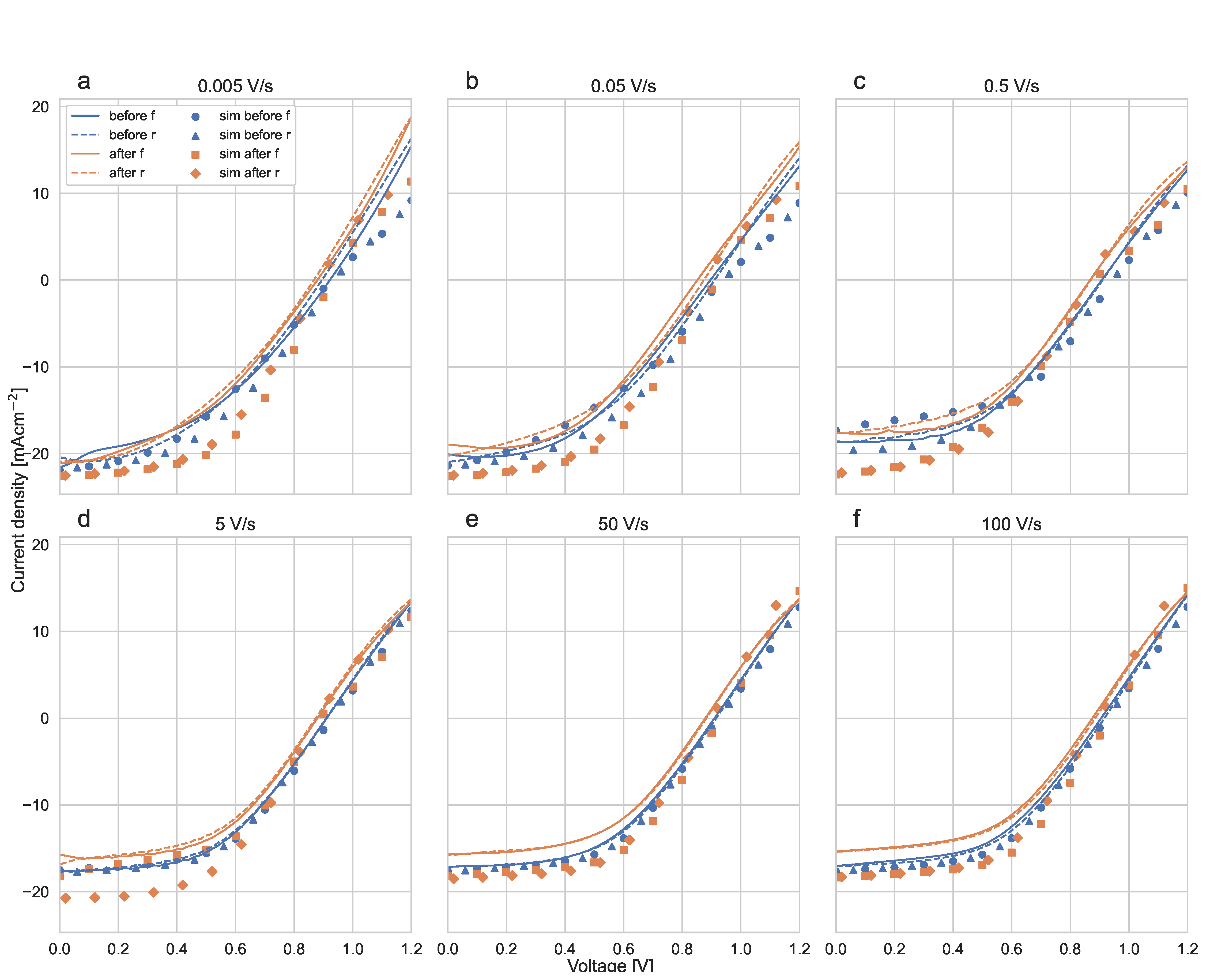}
    \caption{Measured J-V curves with Paios and simulation based on AE estimated parameters for the device kept at $\rm V_{OC}$ before and after stressing.}
    \label{fig:paios_114}
\end{figure}

The parameter estimates obtained from PAIOS measurements, listed in Table \ref{tab:paiosparams}, where the devices were illuminated under a calibrated solar simulator, show noticeable deviations from those extracted during device stressing under white LED illumination. When these PAIOS-derived parameters are used in the DD simulations, the simulated J-V characteristics exhibit a significantly better agreement with the experimentally measured J-V curves (Figures \ref{fig:paios_112} \& \ref{fig:paios_114}).
The discrepancies between the two sets of measurements likely originate from differences in the illumination spectra and associated photo-generation profiles, and the consequences associated with this, as well as from variations in contact resistance, and parasitic capacitances between the two setups. The model training was done with the simulations which used the solar spectra, hence the parameters extracted from the PAIOS measurements appear physically more consistent and yield quantitatively accurate reproductions of the J-V behaviour. Therefore, these parameters are considered more reliable.

Although the drift–diffusion simulations used to train the AE are based on a one-dimensional approximation, the experimental device employs a mesoporous scaffold, where recombination and transport occur over an extended interfacial area. The increased apparent $S$ predicted by the AE may therefore represent an effective parameter capturing both interfacial degradation and geometric enhancement of recombination pathways inherent to the porous structure.

The parameter estimates can be used as a starting point to refine device simulation, which can contribute to a deeper understanding of a device and push the technology further to more stable and efficient devices.

\bibliographystyle{IEEEtran}
\bibliography{bibliography.bib}

@article{odabasi,
title = {Machine learning analysis on stability of perovskite solar cells},
journal = {Solar Energy Materials and Solar Cells},
volume = {205},
pages = {110284},
year = {2020},
issn = {0927-0248},
doi = {https://doi.org/10.1016/j.solmat.2019.110284},
url = {https://www.sciencedirect.com/science/article/pii/S0927024819306130},
author = {Çağla Odabaşı and Ramazan Yıldırım},
}

@article{graniero,
author={Graniero, Paolo  and Khenkin, Mark  and Köbler, Hans  and Hartono, Noor Titan Putri  and Schlatmann, Rutger  and Abate, Antonio  and Unger, Eva  and Jacobsson, T. Jesper  and Ulbrich, Carolin },        
title={The challenge of studying perovskite solar cells’ stability with machine learning},       
journal={Frontiers in Energy Research},        
volume={Volume 11 - 2023},
year={2023},
url={https://www.frontiersin.org/journals/energy-research/articles/10.3389/fenrg.2023.1118654},
doi={10.3389/fenrg.2023.1118654},
issn={2296-598X},
}

@article{hartono,
	title = {Stability follows efficiency based on the analysis of a large perovskite solar cells ageing dataset},
	volume = {14},
	issn = {2041-1723},
	url = {https://doi.org/10.1038/s41467-023-40585-3},
	doi = {10.1038/s41467-023-40585-3},
	number = {1},
	journal = {Nature Communications},
	author = {Hartono, Noor Titan Putri and Köbler, Hans and Graniero, Paolo and Khenkin, Mark and Schlatmann, Rutger and Ulbrich, Carolin and Abate, Antonio},
	month = aug,
	year = {2023},
	pages = {4869},
}

@article{kouroudis,
	title = {Artificial {Intelligence}-{Based}, {Wavelet}-{Aided} {Prediction} of {Long}-{Term} {Outdoor} {Performance} of {Perovskite} {Solar} {Cells}},
	volume = {9},
	url = {https://doi.org/10.1021/acsenergylett.4c00328},
	doi = {10.1021/acsenergylett.4c00328},
	number = {4},
	journal = {ACS Energy Letters},
	author = {Kouroudis, Ioannis and Tanko, Kenedy Tabah and Karimipour, Masoud and Ali, Aziz Ben and Kumar, D. Kishore and Sudhakar, Vediappan and Gupta, Ritesh Kant and Visoly-Fisher, Iris and Lira-Cantu, Monica and Gagliardi, Alessio},
	month = apr,
	year = {2024},
	note = {Publisher: American Chemical Society},
	pages = {1581--1586},
}

@article{zhao,
	title = {Discovery of temperature-induced stability reversal in perovskites using high-throughput robotic learning},
	volume = {12},
	issn = {2041-1723},
	url = {https://doi.org/10.1038/s41467-021-22472-x},
	doi = {10.1038/s41467-021-22472-x},
	number = {1},
	journal = {Nature Communications},
	author = {Zhao, Yicheng and Zhang, Jiyun and Xu, Zhengwei and Sun, Shijing and Langner, Stefan and Hartono, Noor Titan Putri and Heumueller, Thomas and Hou, Yi and Elia, Jack and Li, Ning and Matt, Gebhard J. and Du, Xiaoyan and Meng, Wei and Osvet, Andres and Zhang, Kaicheng and Stubhan, Tobias and Feng, Yexin and Hauch, Jens and Sargent, Edward H. and Buonassisi, Tonio and Brabec, Christoph J.},
	month = apr,
	year = {2021},
	pages = {2191},
}

@article{mammeri,
	title = {Stability forecasting of perovskite solar cells utilizing various machine learning and deep learning techniques},
	issn = {0974-6900},
	url = {https://doi.org/10.1007/s12596-024-01819-9},
	doi = {10.1007/s12596-024-01819-9},journal = {Journal of Optics},
	author = {Mammeri, M. and Bencherif, H. and Dehimi, L. and Hajri, A. and Sasikumar, P. and Syed, A. and AL-Shwaiman, Hind A.},
	month = apr,
	year = {2024},
}

@article{dunlap,
author ="Dunlap-Shohl, Wiley A. and Meng, Yuhuan and Sunkari, Preetham P. and Beck, David A. C. and Meilă, Marina and Hillhouse, Hugh W.",
title  ="Physiochemical machine learning models predict operational lifetimes of CH3NH3PbI3 perovskite solar cells",
journal  ="J. Mater. Chem. A",
year  ="2024",
volume  ="12",
issue  ="16",
pages  ="9730-9746",
publisher  ="The Royal Society of Chemistry",
doi  ="10.1039/D3TA06668A",
url  ="http://dx.doi.org/10.1039/D3TA06668A",
}

@article{nie,
	title = {Light-activated photocurrent degradation and self-healing in perovskite solar cells},
	volume = {7},
	issn = {2041-1723},
	url = {https://doi.org/10.1038/ncomms11574},
	doi = {10.1038/ncomms11574},
	number = {1},
	journal = {Nature Communications},
	author = {Nie, Wanyi and Blancon, Jean-Christophe and Neukirch, Amanda J. and Appavoo, Kannatassen and Tsai, Hsinhan and Chhowalla, Manish and Alam, Muhammad A. and Sfeir, Matthew Y. and Katan, Claudine and Even, Jacky and Tretiak, Sergei and Crochet, Jared J. and Gupta, Gautam and Mohite, Aditya D.},
	month = may,
	year = {2016},
}

@article{domanski,
	title = {Systematic investigation of the impact of operation conditions on the degradation behaviour of perovskite solar cells},
	volume = {3},
	issn = {2058-7546},
	url = {https://doi.org/10.1038/s41560-017-0060-5},
	doi = {10.1038/s41560-017-0060-5},
number = {1},
	journal = {Nature Energy},
	author = {Domanski, Konrad and Alharbi, Essa A. and Hagfeldt, Anders and Grätzel, Michael and Tress, Wolfgang},
	month = jan,
	year = {2018},
	pages = {61--67},
}

@article{kojima,
	title = {Organometal {Halide} {Perovskites} as {Visible}-{Light} {Sensitizers} for {Photovoltaic} {Cells}},
	volume = {131},
	issn = {0002-7863},
	url = {https://doi.org/10.1021/ja809598r},
	doi = {10.1021/ja809598r},
	number = {17},
	journal = {Journal of the American Chemical Society},
	author = {Kojima, Akihiro and Teshima, Kenjiro and Shirai, Yasuo and Miyasaka, Tsutomu},
	month = may,
	year = {2009},
	note = {Publisher: American Chemical Society},
	pages = {6050--6051},
}

@article{leguy,
	title = {Reversible {Hydration} of {CH3NH3PbI3} in {Films}, {Single} {Crystals}, and {Solar} {Cells}},
	volume = {27},
	issn = {0897-4756},
	url = {https://doi.org/10.1021/acs.chemmater.5b00660},
	doi = {10.1021/acs.chemmater.5b00660},
	number = {9},
	journal = {Chemistry of Materials},
	author = {Leguy, Aurélien M. A. and Hu, Yinghong and Campoy-Quiles, Mariano and Alonso, M. Isabel and Weber, Oliver J. and Azarhoosh, Pooya and van Schilfgaarde, Mark and Weller, Mark T. and Bein, Thomas and Nelson, Jenny and Docampo, Pablo and Barnes, Piers R. F.},
	month = may,
	year = {2015},
	note = {Publisher: American Chemical Society},
	pages = {3397--3407},
}

@article{dongxu,
author = {Lin, Dongxu and Gao, Yujia and Zhang, Tiankai and Zhan, Zhenye and Pang, Nana and Wu, Zongwang and Chen, Ke and Shi, Tingting and Pan, Zhenqiang and Liu, Pengyi and Xie, Weiguang},
title = {Vapor Deposited Pure a-FAPbI3 Perovskite Solar Cell via Moisture-Induced Phase Transition Strategy},
journal = {Advanced Functional Materials},
volume = {32},
number = {48},
pages = {2208392},
doi = {https://doi.org/10.1002/adfm.202208392},
url = {https://advanced.onlinelibrary.wiley.com/doi/abs/10.1002/adfm.202208392},
eprint = {https://advanced.onlinelibrary.wiley.com/doi/pdf/10.1002/adfm.202208392},
year = {2022}
}

@article{bryant,
author ="Bryant, Daniel and Aristidou, Nicholas and Pont, Sebastian and Sanchez-Molina, Irene and Chotchunangatchaval, Thana and Wheeler, Scot and Durrant, James R. and Haque, Saif A.",
title  ="Light and oxygen induced degradation limits the operational stability of methylammonium lead triiodide perovskite solar cells",
journal  ="Energy Environ. Sci.",
year  ="2016",
volume  ="9",
issue  ="5",
pages  ="1655-1660",
publisher  ="The Royal Society of Chemistry",
doi  ="10.1039/C6EE00409A",
url  ="http://dx.doi.org/10.1039/C6EE00409A",
}

@article{aristidou,
author = {Aristidou, Nicholas and Sanchez-Molina, Irene and Chotchuangchutchaval, Thana and Brown, Michael and Martinez, Luis and Rath, Thomas and Haque, Saif A.},
title = {The Role of Oxygen in the Degradation of Methylammonium Lead Trihalide Perovskite Photoactive Layers},
journal = {Angewandte Chemie International Edition},
volume = {54},
number = {28},
pages = {8208-8212},
doi = {https://doi.org/10.1002/anie.201503153},
url = {https://onlinelibrary.wiley.com/doi/abs/10.1002/anie.201503153},
eprint = {https://onlinelibrary.wiley.com/doi/pdf/10.1002/anie.201503153},
year = {2015}
}

@article{jacobs,
author ="Jacobs, Daniel A. and Wolff, Christian M. and Chin, Xin-Yu and Artuk, Kerem and Ballif, Christophe and Jeangros, Quentin",
title  ="Lateral ion migration accelerates degradation in halide perovskite devices",
journal  ="Energy Environ. Sci.",
year  ="2022",
volume  ="15",
issue  ="12",
pages  ="5324-5339",
publisher  ="The Royal Society of Chemistry",
doi  ="10.1039/D2EE02330J",
url  ="http://dx.doi.org/10.1039/D2EE02330J",
}

@article{girolamo,
author = {Di Girolamo, Diego and Phung, Nga and Kosasih, Felix Utama and Di Giacomo, Francesco and Matteocci, Fabio and Smith, Joel A. and Flatken, Marion A. and Köbler, Hans and Turren Cruz, Silver H. and Mattoni, Alessandro and Cinà, Lucio and Rech, Bernd and Latini, Alessandro and Divitini, Giorgio and Ducati, Caterina and Di Carlo, Aldo and Dini, Danilo and Abate, Antonio},
title = {Ion Migration-Induced Amorphization and Phase Segregation as a Degradation Mechanism in Planar Perovskite Solar Cells},
journal = {Advanced Energy Materials},
volume = {10},
number = {25},
pages = {2000310},
doi = {https://doi.org/10.1002/aenm.202000310},
url = {https://advanced.onlinelibrary.wiley.com/doi/abs/10.1002/aenm.202000310},
eprint = {https://advanced.onlinelibrary.wiley.com/doi/pdf/10.1002/aenm.202000310},
year = {2020}
}

@article{li_light,
	title = {Light-{Induced} {Degradation} of {CH3NH3PbI3} {Hybrid} {Perovskite} {Thin} {Film}},
	volume = {121},
	issn = {1932-7447},
	url = {https://doi.org/10.1021/acs.jpcc.6b11853},
	doi = {10.1021/acs.jpcc.6b11853},
	number = {7},
	journal = {The Journal of Physical Chemistry C},
	author = {Li, Youzhen and Xu, Xuemei and Wang, Congcong and Ecker, Ben and Yang, Junliang and Huang, Jinsong and Gao, Yongli},
	month = feb,
	year = {2017},
	note = {Publisher: American Chemical Society},
	pages = {3904--3910},
}

@article{min-cheol,
author ="Kim, Min-cheol and Ahn, Namyoung and Lim, Eunhak and Jin, Young Un and Pikhitsa, Peter V. and Heo, Jiyoung and Kim, Seong Keun and Jung, Hyun Suk and Choi, Mansoo",
title  ="Degradation of CH3NH3PbI3 perovskite materials by localized charges and its polarity dependency",
journal  ="J. Mater. Chem. A",
year  ="2019",
volume  ="7",
issue  ="19",
pages  ="12075-12085",
publisher  ="The Royal Society of Chemistry",
doi  ="10.1039/C9TA03180D",
url  ="http://dx.doi.org/10.1039/C9TA03180D",
}

@article{conings,
author = {Conings, Bert and Drijkoningen, Jeroen and Gauquelin, Nicolas and Babayigit, Aslihan and D'Haen, Jan and D'Olieslaeger, Lien and Ethirajan, Anitha and Verbeeck, Jo and Manca, Jean and Mosconi, Edoardo and Angelis, Filippo De and Boyen, Hans-Gerd},
title = {Intrinsic Thermal Instability of Methylammonium Lead Trihalide Perovskite},
journal = {Advanced Energy Materials},
volume = {5},
number = {15},
pages = {1500477},
doi = {https://doi.org/10.1002/aenm.201500477},
url = {https://advanced.onlinelibrary.wiley.com/doi/abs/10.1002/aenm.201500477},
eprint = {https://advanced.onlinelibrary.wiley.com/doi/pdf/10.1002/aenm.201500477},
year = {2015}
}

@article{sominpark,
author = {So Min Park  and Mingyang Wei  and Jian Xu  and Harindi R. Atapattu  and Felix T. Eickemeyer  and Kasra Darabi  and Luke Grater  and Yi Yang  and Cheng Liu  and Sam Teale  and Bin Chen  and Hao Chen  and Tonghui Wang  and Lewei Zeng  and Aidan Maxwell  and Zaiwei Wang  and Keerthan R. Rao  and Zhuoyun Cai  and Shaik M. Zakeeruddin  and Jonathan T. Pham  and Chad M. Risko  and Aram Amassian  and Mercouri G. Kanatzidis  and Kenneth R. Graham  and Michael Grätzel  and Edward H. Sargent },
title = {Engineering ligand reactivity enables high-temperature operation of stable perovskite solar cells},
journal = {Science},
volume = {381},
number = {6654},
pages = {209-215},
year = {2023},
doi = {10.1126/science.adi4107},
URL = {https://www.science.org/doi/abs/10.1126/science.adi4107},
eprint = {https://www.science.org/doi/pdf/10.1126/science.adi4107},
}

@article{seongkang,
author ="Kang, Seong Min and Ahn, Namyoung and Lee, Jin-Wook and Choi, Mansoo and Park, Nam-Gyu",
title  ="Water-repellent perovskite solar cell",
journal  ="J. Mater. Chem. A",
year  ="2014",
volume  ="2",
issue  ="47",
pages  ="20017-20021",
publisher  ="The Royal Society of Chemistry",
doi  ="10.1039/C4TA05413J",
url  ="http://dx.doi.org/10.1039/C4TA05413J",
}

@article{jaesung,
author = {Yun, Jae Sung and Kim, Jincheol and Young, Trevor and Patterson, Robert J. and Kim, Dohyung and Seidel, Jan and Lim, Sean and Green, Martin A. and Huang, Shujuan and Ho-Baillie, Anita},
title = {Humidity-Induced Degradation via Grain Boundaries of HC(NH2)2PbI3 Planar Perovskite Solar Cells},
journal = {Advanced Functional Materials},
volume = {28},
number = {11},
pages = {1705363},
doi = {https://doi.org/10.1002/adfm.201705363},
url = {https://advanced.onlinelibrary.wiley.com/doi/abs/10.1002/adfm.201705363},
eprint = {https://advanced.onlinelibrary.wiley.com/doi/pdf/10.1002/adfm.201705363},
year = {2018}
}

@article{grancini,
	title = {One-{Year} stable perovskite solar cells by {2D}/{3D} interface engineering},
	volume = {8},
	issn = {2041-1723},
	url = {https://doi.org/10.1038/ncomms15684},
	doi = {10.1038/ncomms15684},
	number = {1},
	journal = {Nature Communications},
	author = {Grancini, G. and Roldán-Carmona, C. and Zimmermann, I. and Mosconi, E. and Lee, X. and Martineau, D. and Narbey, S. and Oswald, F. and De Angelis, F. and Graetzel, M. and Nazeeruddin, Mohammad Khaja},
	month = jun,
	year = {2017},
	pages = {15684},
}

@article{boyd,
	title = {Barrier {Design} to {Prevent} {Metal}-{Induced} {Degradation} and {Improve} {Thermal} {Stability} in {Perovskite} {Solar} {Cells}},
	volume = {3},
	url = {https://doi.org/10.1021/acsenergylett.8b00926},
	doi = {10.1021/acsenergylett.8b00926},
	number = {7},
	journal = {ACS Energy Letters},
	author = {Boyd, Caleb C. and Cheacharoen, Rongrong and Bush, Kevin A. and Prasanna, Rohit and Leijtens, Tomas and McGehee, Michael D.},
	month = jul,
	year = {2018},
	note = {Publisher: American Chemical Society},
	pages = {1772--1778},
}

@article{zhaojie,
author ="Zhao, Jie and Rai, Nitish and Benitez-Rodriguez, Juan F. and Yan, Wenqi and Sutherland, Luke J. and Yan, Junlin and Chesman, Anthony S. R. and Jasieniak, Jacek and Lu, Jianfeng and Bach, Udo",
title  ="Cation optimization for bifacial surface passivation in efficient and stable perovskite solar cells",
journal  ="EES Sol.",
year  ="2025",
volume  ="1",
issue  ="1",
pages  ="56-65",
publisher  ="RSC",
doi  ="10.1039/D4EL00013G",
url  ="http://dx.doi.org/10.1039/D4EL00013G",
}

@misc{nrel,
title={Best {R}esearch-{C}ell {E}fficiency {C}hart plotted by {N}ational {R}enewable {E}nergy {L}aboratory, {USA}},
howpublished={\url{https://www.nrel.gov/pv/interactive-cell-efficiency.html}, visited on 23.04.2025},
author = {NREL},
lastvisited = {23.04.2025},
}

@misc{solaronix,
author={{Solaronix SA}},
howpublished={\url{https://www.solaronix.com/}, visited on 12.12.2024},
lastvisited = {12.12.2024}
}

@article{carbonrecord,
	title = {Enhancing hole-conductor-free, printable mesoscopic perovskite solar cells through post-fabrication treatment via electrophilic reaction},
	issn = {2058-7546},
	url = {https://doi.org/10.1038/s41560-025-01823-8},
	doi = {10.1038/s41560-025-01823-8},
	journal = {Nature Energy},
	author = {Ma, Yongming and Liu, Jiale and Chen, Xiayan and Zhao, Xinran and Qi, Jianhang and She, Bin and Liu, Shuang and Jiang, Youyu and Sheng, Yusong and Han, Chuanzhou and Zhang, Guodong and Xie, Jiayu and Chen, Kai and Cheng, Yanjie and Xiang, Junwei and Yang, Li-Ming and Zhou, Yang and Ling, Furi and Zhou, Yinhua and Mei, Anyi and Han, Hongwei},
	month = aug,
	year = {2025},
}

@article{bogachuk_comparison_2021,
	title = {Comparison of highly conductive natural and synthetic graphites for electrodes in perovskite solar cells},
	volume = {178},
	issn = {0008-6223},
	url = {https://www.sciencedirect.com/science/article/pii/S0008622321000300},
	doi = {https://doi.org/10.1016/j.carbon.2021.01.022},
	journal = {Carbon},
	author = {Bogachuk, Dmitry and Tsuji, Ryuki and Martineau, David and Narbey, Stephanie and Herterich, Jan P. and Wagner, Lukas and Suginuma, Kumiko and Ito, Seigo and Hinsch, Andreas},
	year = {2021},
	pages = {10--18},
}

@article{kerremans,
author = {Kerremans, Robin and Sandberg, Oskar J. and Meroni, Simone and Watson, Trystan and Armin, Ardalan and Meredith, Paul},
title = {On the Electro-Optics of Carbon Stack Perovskite Solar Cells},
journal = {Solar RRL},
volume = {4},
number = {2},
pages = {1900221},
doi = {https://doi.org/10.1002/solr.201900221},
url = {https://onlinelibrary.wiley.com/doi/abs/10.1002/solr.201900221},
year = {2020}
}

@article{neukom_consistent_2019,
	title = {Consistent {Device} {Simulation} {Model} {Describing} {Perovskite} {Solar} {Cells} in {Steady}-{State}, {Transient}, and {Frequency} {Domain}},
	volume = {11},
	issn = {1944-8244, 1944-8252},
	url = {https://pubs.acs.org/doi/10.1021/acsami.9b04991},
	doi = {10.1021/acsami.9b04991},
	language = {en},
	number = {26},
	journal = {ACS Applied Materials \& Interfaces},
	author = {Neukom, Martin T. and Schiller, Andreas and Züfle, Simon and Knapp, Evelyne and Ávila, Jorge and Pérez-del-Rey, Daniel and Dreessen, Chris and Zanoni, Kassio P.S. and Sessolo, Michele and Bolink, Henk J. and Ruhstaller, Beat},
	month = jul,
	year = {2019},
	pages = {23320--23328}
}

@article{trpltpctpvtau,
author = {Wang, Liang and Miao, Qingqing and Wang, Dandan and Chen, Mengmeng and Bi, Huan and Liu, Jiaqi and Baranwal, Ajay Kumar and Kapil, Gaurav and Sanehira, Yoshitaka and Kitamura, Takeshi and Ma, Tingli and Zhang, Zheng and Shen, Qing and Hayase, Shuzi},
title = {14.31 \% Power Conversion Efficiency of Sn-Based Perovskite Solar Cells via Efficient Reduction of Sn4+},
journal = {Angewandte Chemie International Edition},
volume = {62},
number = {33},
pages = {e202307228},
doi = {https://doi.org/10.1002/anie.202307228},
url = {https://onlinelibrary.wiley.com/doi/abs/10.1002/anie.202307228},
eprint = {https://onlinelibrary.wiley.com/doi/pdf/10.1002/anie.202307228},
year = {2023}
}

@article{meins_1,
author = {Zbinden, Oliver and Knapp, Evelyne and Tress, Wolfgang},
title = {Identifying Performance Limiting Parameters in Perovskite Solar Cells Using Machine Learning},
journal = {Solar RRL},
volume = {8},
number = {6},
pages = {2300999},
doi = {https://doi.org/10.1002/solr.202300999},
url = {https://onlinelibrary.wiley.com/doi/abs/10.1002/solr.202300999},
eprint = {https://onlinelibrary.wiley.com/doi/pdf/10.1002/solr.202300999},
year = {2024}
}

@article{thesejv,
author = {These, Albert and Koster, L. Jan Anton and Brabec, Christoph J. and Le Corre, Vincent M.},
title = {Beginner's Guide to Visual Analysis of Perovskite and Organic Solar Cell Current Density–Voltage Characteristics},
journal = {Advanced Energy Materials},
volume = {14},
number = {21},
pages = {2400055},
doi = {https://doi.org/10.1002/aenm.202400055},
url = {https://advanced.onlinelibrary.wiley.com/doi/abs/10.1002/aenm.202400055},
eprint = {https://advanced.onlinelibrary.wiley.com/doi/pdf/10.1002/aenm.202400055},
year = {2024}
}

@article{zhang_degradation_2022,
	title = {Degradation pathways in perovskite solar cells and how to meet international standards},
	volume = {3},
	issn = {2662-4443},
	url = {https://doi.org/10.1038/s43246-022-00281-z},
	doi = {10.1038/s43246-022-00281-z},
	number = {1},
	journal = {Communications Materials},
	author = {Zhang, Deyi and Li, Daiyu and Hu, Yue and Mei, Anyi and Han, Hongwei},
	month = aug,
	year = {2022},
	pages = {58},
}

@article{kim_degrad_chargemob,
author = {Kim, Souk Y. and Kumachang, Cyril C. F. and Doumon, Nutifafa Y.},
title = {Characterization Tools to Probe Degradation Mechanisms in Organic and Perovskite Solar Cells},
journal = {Solar RRL},
volume = {7},
number = {13},
pages = {2300155},
doi = {https://doi.org/10.1002/solr.202300155},
url = {https://onlinelibrary.wiley.com/doi/abs/10.1002/solr.202300155},
eprint = {https://onlinelibrary.wiley.com/doi/pdf/10.1002/solr.202300155},
year = {2023}
}

@article{dunfield_deg_incr_ion,
author = {Dunfield, Sean P. and Bliss, Lyle and Zhang, Fei and Luther, Joseph M. and Zhu, Kai and van Hest, Maikel F. A. M. and Reese, Matthew O. and Berry, Joseph J.},
title = {From Defects to Degradation: A Mechanistic Understanding of Degradation in Perovskite Solar Cell Devices and Modules},
journal = {Advanced Energy Materials},
volume = {10},
number = {26},
pages = {1904054},
doi = {https://doi.org/10.1002/aenm.201904054},
url = {https://advanced.onlinelibrary.wiley.com/doi/abs/10.1002/aenm.201904054},
eprint = {https://advanced.onlinelibrary.wiley.com/doi/pdf/10.1002/aenm.201904054},
year = {2020}
}

@article{hyster_tress_1,
author = {Tress, Wolfgang and Correa Baena, Juan Pablo and Saliba, Michael and Abate, Antonio and Graetzel, Michael},
title = {Inverted Current–Voltage Hysteresis in Mixed Perovskite Solar Cells: Polarization, Energy Barriers, and Defect Recombination},
journal = {Advanced Energy Materials},
volume = {6},
number = {19},
pages = {1600396},
doi = {https://doi.org/10.1002/aenm.201600396},
url = {https://advanced.onlinelibrary.wiley.com/doi/abs/10.1002/aenm.201600396},
eprint = {https://advanced.onlinelibrary.wiley.com/doi/pdf/10.1002/aenm.201600396},
year = {2016}
}

@article{richardson_hyster,
author ="Richardson, Giles and O{'}Kane, Simon E. J. and Niemann, Ralf G. and Peltola, Timo A. and Foster, Jamie M. and Cameron, Petra J. and Walker, Alison B.",
title  ="Can slow-moving ions explain hysteresis in the current–voltage curves of perovskite solar cells?",
journal  ="Energy Environ. Sci.",
year  ="2016",
volume  ="9",
issue  ="4",
pages  ="1476-1485",
publisher  ="The Royal Society of Chemistry",
doi  ="10.1039/C5EE02740C",
url  ="http://dx.doi.org/10.1039/C5EE02740C",
}

@article{tress_metal_2017,
	title = {Metal {Halide} {Perovskites} as {Mixed} {Electronic}–{Ionic} {Conductors}: {Challenges} and {Opportunities}—{From} {Hysteresis} to {Memristivity}},
	volume = {8},
	url = {https://doi.org/10.1021/acs.jpclett.7b00975},
	doi = {10.1021/acs.jpclett.7b00975},
	number = {13},
	journal = {The Journal of Physical Chemistry Letters},
	author = {Tress, Wolfgang},
	month = jul,
	year = {2017},
	note = {Publisher: American Chemical Society},
	pages = {3106--3114},
}

@article{https://doi.org/10.1002/aenm.202403850,
author = {Torre Cachafeiro, Miguel A. and Comi, Ennio Luigi and Parayil Shaji, Sharun and Narbey, Stéphanie and Jenatsch, Sandra and Knapp, Evelyne and Tress, Wolfgang},
title = {Ion Migration in Mesoscopic Perovskite Solar Cells: Effects on Electroluminescence, Open Circuit Voltage, and Photovoltaic Quantum Efficiency},
journal = {Advanced Energy Materials},
volume = {15},
number = {5},
pages = {2403850},
doi = {https://doi.org/10.1002/aenm.202403850},
url = {https://advanced.onlinelibrary.wiley.com/doi/abs/10.1002/aenm.202403850},
eprint = {https://advanced.onlinelibrary.wiley.com/doi/pdf/10.1002/aenm.202403850},
year = {2025}
}

@article{thiesbrummel_ion_induced_2024,
	title = {Ion-induced field screening as a dominant factor in perovskite solar cell operational stability},
	volume = {9},
	issn = {2058-7546},
	url = {https://doi.org/10.1038/s41560-024-01487-w},
	doi = {10.1038/s41560-024-01487-w},
	number = {6},
	journal = {Nature Energy},
	author = {Thiesbrummel, Jarla and Shah, Sahil and Gutierrez-Partida, Emilio and Zu, Fengshuo and Peña-Camargo, Francisco and Zeiske, Stefan and Diekmann, Jonas and Ye, Fangyuan and Peters, Karol P. and Brinkmann, Kai O. and Caprioglio, Pietro and Dasgupta, Akash and Seo, Seongrok and Adeleye, Fatai A. and Warby, Jonathan and Jeangros, Quentin and Lang, Felix and Zhang, Shuo and Albrecht, Steve and Riedl, Thomas and Armin, Ardalan and Neher, Dieter and Koch, Norbert and Wu, Yongzhen and Le Corre, Vincent M. and Snaith, Henry and Stolterfoht, Martin},
	month = jun,
	year = {2024},
	pages = {664--676},
}

@article{CHE20171064,
title = {F-doped TiO2 Compact Film for High-Efficient Perovskite Solar Cells},
journal = {International Journal of Electrochemical Science},
volume = {12},
number = {2},
pages = {1064-1074},
year = {2017},
issn = {1452-3981},
doi = {https://doi.org/10.20964/2017.02.21},
url = {https://www.sciencedirect.com/science/article/pii/S1452398123105396},
author = {M. Che and Y. Fang and J. Yuan and Y. Zhu and Q. Liu and J. Song}
}

@article{lifetimestuff,
author = {Azzouzi, Mohammed and Calado, Philip and Telford, Andrew M. and Eisner, Flurin and Hou, Xueyan and Kirchartz, Thomas and Barnes, Piers R. F. and Nelson, Jenny},
title = {Overcoming the Limitations of Transient Photovoltage Measurements for Studying Recombination in Organic Solar Cells},
journal = {Solar RRL},
volume = {4},
number = {5},
pages = {1900581},
doi = {https://doi.org/10.1002/solr.201900581},
url = {https://onlinelibrary.wiley.com/doi/abs/10.1002/solr.201900581},
eprint = {https://onlinelibrary.wiley.com/doi/pdf/10.1002/solr.201900581},
year = {2020}
}

@article{bisquert2021frequency,
  title={From frequency domain to time transient methods for halide perovskite solar cells: The connections of IMPS, IMVS, TPC, and TPV},
  author={Bisquert, Juan and Janssen, Mathijs},
  journal={The Journal of Physical Chemistry Letters},
  volume={12},
  number={33},
  pages={7964--7971},
  year={2021},
  publisher={ACS Publications}
}

@article{PhysRevApplied.11.044005,
  title = {Identifying Dominant Recombination Mechanisms in Perovskite Solar Cells by Measuring the Transient Ideality Factor},
  author = {Calado, Phil and Burkitt, Dan and Yao, Jizhong and Troughton, Joel and Watson, Trystan M. and Carnie, Matt J. and Telford, Andrew M. and O'Regan, Brian C. and Nelson, Jenny and Barnes, Piers R.F.},
  journal = {Phys. Rev. Appl.},
  volume = {11},
  issue = {4},
  pages = {044005},
  numpages = {17},
  year = {2019},
  month = {Apr},
  publisher = {American Physical Society},
  doi = {10.1103/PhysRevApplied.11.044005},
  url = {https://link.aps.org/doi/10.1103/PhysRevApplied.11.044005}
}

@article{AE_paper,
  title = {Autoencoder for Parameter Estimation and Current-Voltage Curve Simulation of Perovskite Solar Cells},
  url = {http://dx.doi.org/10.21203/rs.3.rs-8037624/v1},
  DOI = {10.21203/rs.3.rs-8037624/v1},
  publisher = {Springer Science and Business Media LLC},
  author = {Zbinden,  Oliver and Comi,  Ennio Luigi and Knapp,  Evelyne and Tress,  Wolfgang},
  year = {2025},
journal = {Research Square (preprint)}
}

@article{plateau,
  title={Impedance spectroscopy of metal halide perovskite solar cells from the perspective of equivalent circuits},
  author={Guerrero, Antonio and Bisquert, Juan and Garcia-Belmonte, Germa},
  journal={Chemical Reviews},
  volume={121},
  number={23},
  pages={14430--14484},
  year={2021},
  publisher={ACS Publications}
}

\clearpage
\onecolumn
\section{Supporting information}

%\begin{figure}[h]
%    \centering
%    \includegraphics[width=.25\linewidth]{img/stack.png}
%    \caption{Illustration of the device stack of the mesoporous PSCs used in experiment and simulation.}
%    \label{fig:devstackdegr}
%\end{figure}

\begin{figure}[t]
    \centering
    \includegraphics[width=1\linewidth]{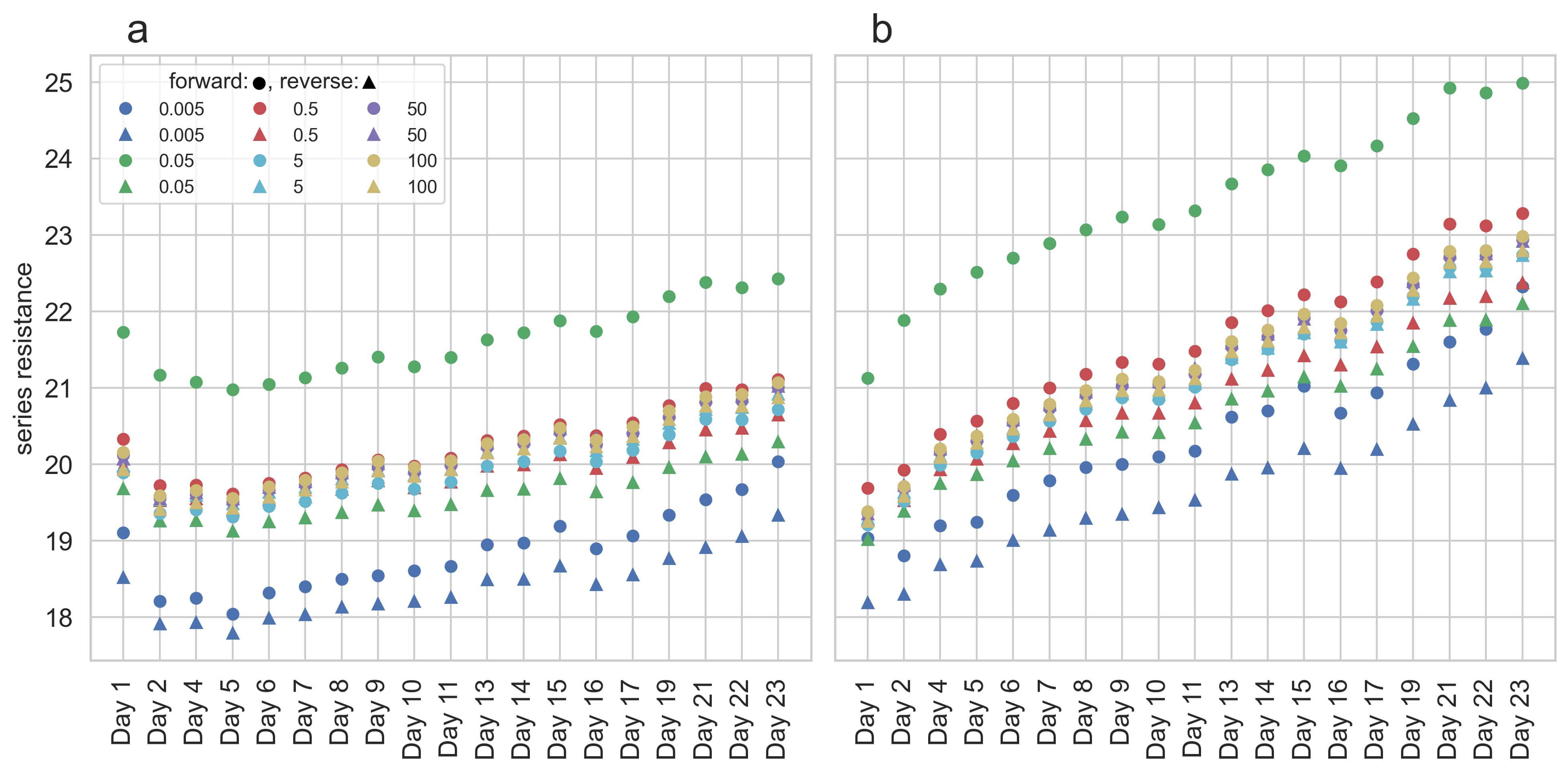}
    \caption{Differential series resistance between $0.8$ V and $1$ V for a: device kept at MPP, and b: device kept at $\rm V_{OC}$, for each day, scan speed, and scanning direction.}
    \label{fig:seriesresistance}
\end{figure}

\begin{table*}[t]
    \centering
    \caption{Complementary to Figure \ref{fig:param_est_secondstudy}, the numeric values are listed in this table for both devices}
    \label{tab:encodedparams}
        \begin{adjustbox}{width=1\textwidth}
        \begin{small}
        \begin{sc}
        \begin{tabular}[t]{llrrrrrr}

        \toprule
        & & $\rm S$ $\rm [cms^{-1}]$ & $\rm \mu$ $\rm [cm^2V^{-1}s^{-1}]$ & $\tau$ $\rm [ns]$ & $\mu_{TiO_2}$ $\rm [cm^2V^{-1}s^{-1}]$ & $\rm\mu_C$ $\rm [cm^2V^{-1}s^{-1}]$ & $\rm\rho_{Ion}$ $\rm [cm^{-3}]$ \\
         
        \midrule

        MPP-tracked & Day 1 & $23.09$ & $17.51$ & $3.14$ & $7.81\cdot10^{-5}$ & $2.16\cdot10^{-8}$ & $2.51\cdot10^{17}$ \\

        & Day 2 & $20.73$ & $21.76$ & $2.45$ & $8.16\cdot10^{-5}$ & $4.06\cdot10^{-8}$ & $2.45\cdot10^{17}$ \\

        & Day 4 & $23.43$ & $21.39$ & $2.43$ & $8.56\cdot10^{-5}$ & $3.44\cdot10^{-8}$ & $2.47\cdot10^{17}$ \\

        & Day 5 & $25.81$ & $23.28$ & $2.36$ & $9.18\cdot10^{-5}$ & $2.29\cdot10^{-8}$ & $2.35\cdot10^{17}$ \\

        & Day 6 & $24.89$ & $23.31$ & $2.38$ & $8.73\cdot10^{-5}$ & $2.05\cdot10^{-8}$ & $2.41\cdot10^{17}$ \\

        & Day 7 & $26.92$ & $22.11$ & $2.44$ & $9.04\cdot10^{-5}$ & $1.83\cdot10^{-8}$ & $2.52\cdot10^{17}$ \\

        & Day 8 & $30.10$ & $18.31$ & $2.87$ & $9.55\cdot10^{-5}$ & $1.58\cdot10^{-8}$ & $2.62\cdot10^{17}$ \\

        & Day 9 & $29.96$ & $16.96$ & $2.94$ & $9.44\cdot10^{-5}$ & $1.56\cdot10^{-8}$ & $2.59\cdot10^{17}$ \\

        & Day 10 & $33.42$ & $9.97$ & $5.34$ & $9.87\cdot10^{-5}$ & $1.02\cdot10^{-8}$ & $2.36\cdot10^{17}$ \\

        & Day 11 & $34.75$ & $9.41$ & $5.72$ & $9.84\cdot10^{-5}$ & $9.31\cdot10^{-9}$ & $2.58\cdot10^{17}$ \\

        & Day 13 & $36.69$ & $6.50$ & $8.54$ & $9.45\cdot10^{-5}$ & $7.75\cdot10^{-9}$ & $2.75\cdot10^{17}$ \\

        & Day 14 & $39.56$ & $5.15$ & $10.73$ & $9.72\cdot10^{-5}$ & $6.96\cdot10^{-9}$ & $2.73\cdot10^{17}$ \\

        & Day 15 & $40.19$ & $4.14$ & $13.85$ & $9.31\cdot10^{-5}$ & $6.53\cdot10^{-9}$ & $2.89\cdot10^{17}$ \\

        & Day 16 & $49.68$ & $2.53$ & $22.38$ & $1.08\cdot10^{-4}$ & $5.28\cdot10^{-9}$ & $2.61\cdot10^{17}$ \\

        & Day 17 & $49.89$ & $2.29$ & $25.29$ & $1.03\cdot10^{-4}$ & $4.75\cdot10^{-9}$ & $2.77\cdot10^{17}$ \\

        & Day 19 & $48.27$ & $2.10$ & $28.39$ & $9.47\cdot10^{-5}$ & $4.36\cdot10^{-9}$ & $2.94\cdot10^{17}$ \\

        & Day 21 & $48.17$ & $2.37$ & $25.31$ & $9.09\cdot10^{-5}$ & $3.96\cdot10^{-9}$ & $3.12\cdot10^{17}$ \\

        & Day 22 & $52.62$ & $1.10$ & $58.77$ & $8.65\cdot10^{-5}$ & $3.74\cdot10^{-9}$ & $2.87\cdot10^{17}$ \\

        & Day 23 & $48.70$ & $8.55\cdot10^{-1}$ & $77.56$ & $7.84\cdot10^{-5}$ & $4.03\cdot10^{-9}$ & $2.62\cdot10^{17}$ \\

        \midrule

        $\rm V_{OC}$-tracked & Day 1 & $39.68$ & $4.02$ & $21.90$ & $7.65\cdot10^{-5}$ & $1.09\cdot10^{-8}$ & $1.52\cdot10^{17}$ \\

        & Day 2 & $58.05$ & $7.57\cdot10^{-1}$ & $93.61$ & $9.58\cdot10^{-5}$ & $7.19\cdot10^{-9}$ & $1.61\cdot10^{17}$ \\

        & Day 4 & $74.50$ & $1.73\cdot10^{-1}$ & $4.41\cdot10^2$ & $1.03\cdot10^{-4}$ & $4.98\cdot10^{-9}$ & $1.85\cdot10^{17}$ \\

        & Day 5 & $79.51$ & $1.52\cdot10^{-1}$ & $5.42\cdot10^2$ & $1.02\cdot10^{-4}$ & $4.56\cdot10^{-9}$ & $2.03\cdot10^{17}$ \\

        & Day 6 & $79.39$ & $1.34\cdot10^{-1}$ & $6.64\cdot10^2$ & $9.51\cdot10^{-5}$ & $4.71\cdot10^{-9}$ & $2.03\cdot10^{17}$ \\

        & Day 7 & $79.74$ & $1.49\cdot10^{-1}$ & $6.32\cdot10^2$ & $9.06\cdot10^{-5}$ & $4.62\cdot10^{-9}$ & $2.12\cdot10^{17}$ \\

        & Day 8 & $81.70$ & $1.55\cdot10^{-1}$ & $6.24\cdot10^2$ & $8.77\cdot10^{-5}$ & $4.77\cdot10^{-9}$ & $2.01\cdot10^{17}$ \\

        & Day 9 & $86.52$ & $1.59\cdot10^{-1}$ & $6.16\cdot10^2$ & $8.75\cdot10^{-5}$ & $4.62\cdot10^{-9}$ & $2.08\cdot10^{17}$ \\

        & Day 10 & $1.00\cdot10^2$ & $7.44\cdot10^{-2}$ & $6.16\cdot10^2$ & $9.37\cdot10^{-5}$ & $4.72\cdot10^{-9}$ & $2.12\cdot10^{17}$ \\

        & Day 11 & $99.38$ & $8.99\cdot10^{-2}$ & $1.47\cdot10^3$ & $9.00\cdot10^{-5}$ & $4.79\cdot10^{-9}$ & $2.07\cdot10^{17}$ \\

        & Day 13 & $97.10$ & $1.18\cdot10^{-1}$ & $1.12\cdot10^3$ & $8.12\cdot10^{-5}$ & $4.83\cdot10^{-9}$ & $1.97\cdot10^{17}$ \\

        & Day 14 & $95.68$ & $1.45\cdot10^{-1}$ & $8.74\cdot10^2$ & $7.89\cdot10^{-5}$ & $4.95\cdot10^{-9}$ & $1.91\cdot10^{17}$ \\

        & Day 15 & $92.71$ & $1.64\cdot10^{-1}$ & $7.55\cdot10^2$ & $7.38\cdot10^{-5}$ & $5.09\cdot10^{-9}$ & $1.75\cdot10^{17}$ \\

        & Day 16 & $1.20\cdot10^2$ & $9.54\cdot10^{-2}$ & $1.69\cdot10^3$ & $8.55\cdot10^{-5}$ & $4.19\cdot10^{-9}$ & $2.46\cdot10^{17}$ \\

        & Day 17 & $1.15\cdot10^2$ & $1.16\cdot10^{-1}$ & $1.33\cdot10^3$ & $8.20\cdot10^{-5}$ & $4.52\cdot10^{-9}$ & $2.37\cdot10^{17}$ \\

        & Day 19 & $1.07\cdot10^2$ & $1.71\cdot10^{-1}$ & $8.44\cdot10^2$ & $7.61\cdot10^{-5}$ & $4.99\cdot10^{-9}$ & $2.26\cdot10^{17}$ \\

        & Day 21 & $1.03\cdot10^2$ & $2.28\cdot10^{-1}$ & $5.71\cdot10^2$ & $7.24\cdot10^{-5}$ & $4.96\cdot10^{-9}$ & $2.11\cdot10^{17}$ \\

        & Day 22 & $1.16\cdot10^2$ & $1.30\cdot10^{-1}$ & $1.17\cdot10^3$ & $8.70\cdot10^{-5}$ & $6.45\cdot10^{-9}$ & $2.97\cdot10^{17}$ \\

        & Day 23 & $1.08\cdot10^2$ & $1.63\cdot10^{-1}$ & $9.22\cdot10^2$ & $8.34\cdot10^{-5}$ & $8.16\cdot10^{-9}$ & $3.07\cdot10^{17}$ \\

        \bottomrule
    \end{tabular}
    \end{sc}
    \end{small}
    \end{adjustbox}
    \vskip -0.1in
\end{table*}

\end{document}